\documentclass[lettersize,journal]{IEEEtran}
\usepackage{amsmath,amsfonts}
\usepackage{algorithmic}
\usepackage{algorithm}
\usepackage{array}
\usepackage[caption=false,font=normalsize,labelfont=sf,textfont=sf]{subfig}
\usepackage{textcomp}
\usepackage{stfloats}
\usepackage{url}
\usepackage{verbatim}
\usepackage{graphicx}
\usepackage{cite}

\usepackage{booktabs}
\usepackage{xcolor}
\usepackage{multirow}
\usepackage{pifont}
\usepackage{tabularray}
\usepackage{xspace}
\usepackage{bbding}
\newcommand{\myparagraph}[1]{\vspace {3pt}\noindent\textbf{\emph{#1}}}
\newcommand{\sys}{WaferAI-SIM\xspace}
\newenvironment{myitemize}  {\begin{list}{\labelitemi}{\itemsep1pt \topsep2pt \parsep0.00in
  \partopsep=0pt \leftmargin1em}}  {\end{list}}

\begin{document}

\title{Systematic Exploration of Multi-core Architectures for Efficient LLM Serving using WaferAI-SIM}

\author{Tianhao Zhu$^{\dagger}$, Dahu Feng$^{\ast}$, Erhu Feng\textsuperscript{\Envelope}$^{\dagger}$, Yubin Xia$^{\dagger}$, Haibo Chen$^{\dagger}$
        \thanks{$^{\dagger}$Institute of Parallel and Distributed Systems (IPADS), School of Computing, Shanghai Jiao Tong University, Shanghai 200240, China.}
        \thanks{$^{\ast}$Center for Brain-Inspired Computing Research, Optical Memory National Engineering Research Center Tsinghua University--China Electronics Technology HIK Group Co. Joint Research Center for Brain-Inspired Computing, IDG/McGovern Institute for Brain Research at Tsinghua University, Department of Precision Instrument, Tsinghua University, Beijing 100084, China.}
}

\maketitle

\begin{abstract}
With the widespread adoption of Large Language Models (LLMs), the demand for high-performance LLM inference services continues to grow.
Multi-core AI accelerators, such as Groq, Graphcore IPU, and Cerebras WSE, provide promising platforms for LLM serving,
but their distributed memory systems require careful coordination between hardware configuration and serving policies.
Otherwise, mismatched tensor partitioning, data placement, and memory management can substantially underutilize compute and communication resources.

To address these challenges, we present \sys, a multi-level simulation framework that combines transaction-level simulation with an analytical performance model.
\sys enables simulator-driven co-design of LLM serving strategies and multi-core accelerator architectures,
targeting the early design stage in which emerging platforms are not yet broadly available for empirical serving studies.
It captures how LLM serving policies interact with compute-core count, memory hierarchy, and interconnect topology, enabling architecture-aware exploration beyond GPU-centric assumptions.
We evaluate representative LLMs across a range of chip configurations and serving scenarios.
Across the evaluation, \sys reports 1.32$\times$--6.03$\times$ latency improvements,
where the lower endpoint comes from ring-based placement at TP=16 over the placement baselines, and the upper endpoint comes from K-dimension TP over MN-dimension TP for Qwen3\_4B at TP=4 with sequence length 256.
For LLM serving, our findings provide guidance for co-designing hardware architectures and serving strategies
for multi-core AI accelerators across diverse LLM workloads.
\end{abstract}

\begin{IEEEkeywords}
LLM serving, AI accelerator, multi-core NPU, simulator, dataflow architecture, model parallelism, memory management, heterogeneity
\end{IEEEkeywords}

\section{Introduction}
\label{s:intro}

\IEEEPARstart{W}{ith} the rapid advancement of large language models (LLMs)~\cite{bai2023qwen,chatgpt-3.5,openai2023gpt4,Llama,team2023gemini,liu2024deepseek} and the widespread deployment of LLM-powered applications,
such as agents~\cite{Hong_2024_CVPR,huang2024understanding,wang2024mobileagentv2mobiledeviceoperation}, chatbots~\cite{chatgpt-3.5}, code generation~\cite{cursor,githubCopilot}, and autonomous driving~\cite{tesla-automation,chen2024end},
the demand for accelerating LLM inference has garnered substantial attention from both academia and industry.
Therefore, various chip manufacturers have introduced dedicated AI accelerators,
including Graphcore IPUs~\cite{IPU}, Tesla Dojo~\cite{DOJO}, Cerebras WSE~\cite{lie2023cerebras}, and Groq~\cite{Groq}.
These AI accelerators typically feature domain-specific architectures (DSAs) and multi-core designs,
integrating dozens to thousands of compute cores (termed as multi-core AI accelerators in this paper).
Rather than relying on conventional cache-based unified memory architectures (usually adopted in GPUs)~\cite{H100,choquette2020nvidia},
multi-core AI accelerators employ high-speed network-on-chip (NoC)~\cite{NoC-arch} and large per-core scratchpad memory~\cite{prabhakar2022sambanova,chen2016eyeriss,EyerissV2,IPU,SCNN}.
With these architectural innovations, they offer improved scalability, higher performance, and reduced power consumption.
Groq's official report on Anyscale's public LLMPerf leaderboard states that its LPU Inference Engine achieves up to 18$\times$ higher output-token throughput than top cloud-based inference providers for Llama-2-70B~\cite{Groq-faster-than-gpu}.

Although multi-core architectures offer scalable compute resources and high performance, two key challenges remain in efficiently utilizing the available hardware resources.
First, the large and complex hardware design space can lead to substantial resource underutilization.
Second, multi-core architectures substantially complicate distributed data management.
The absence of a unified memory system necessitates careful placement of diverse tensors across cores.
Suboptimal placement of weights and KV caches can incur prohibitive communication overheads and redundant data replication.
Unlike well-studied GPUs,
emerging multi-core and 3D-stacked-memory architectures have limited hardware availability,
making it difficult to address these challenges solely through empirical tuning on deployed serving systems.

To systematically explore optimized hardware configurations for diverse multi-core AI accelerators
(e.g., LPU, wafer-scale accelerators, dataflow accelerators), we introduce \sys, a dedicated simulation framework for the multi-core architecture.
\sys supports flexible configuration of hardware attributes, including compute-core count, memory hierarchy, and interconnect topology, enabling it to model a wide range of multi-core AI accelerator designs.
\sys balances fidelity and efficiency while supporting streaming request inputs for real-world LLM serving workloads.

Leveraging \sys, we conduct a holistic study of on-chip distributed data management and architectural specialization for LLM serving.
Rather than targeting a fixed accelerator instance, we use \sys to evaluate how existing and new serving mechanisms interact with parameterized emerging hardware designs, including wafer-scale and hybrid-memory accelerators.
While prior works~\cite{WaferLLM,T10,9546435,wsc-llm} have explored on-chip tensor parallelism and core placement,
they are often tailored to specific hardware platforms, limiting generality and potentially leading to suboptimal performance.
Building on this capability, we first provide a systematic analysis of tensor partitioning and placement strategies
(e.g., AllReduce vs. AllGather, and mesh vs. ring topologies) across a broad design space.
Second, to address the constraints of discrete memory, we study fine-grained KV cache and weight management strategies
that optimize data movement across the memory hierarchy.
Finally, we explore on-chip heterogeneous architectures for Prefill-Decoding (PD) serving,
evaluating how different hardware configurations adapt to the distinct computational demands of different serving stages.

We evaluate \sys with different serving strategies across a range of representative LLMs, from 1.7B to 235B parameters,
including both dense and MoE models. Experimental results demonstrate that ring-based core placement reaches a 1.32$\times$ endpoint under TP=16 relative to the evaluated placement baselines,
while K-dimension TP partitioning reaches a 6.03$\times$ endpoint over MN-dimension TP for Qwen3\_4B at TP=4 and sequence length 256.
We have open-sourced \sys at \url{https://github.com/IPADS-SAI/WaferAI-SIM} to facilitate future research on multi-core AI accelerator
hardware architecture and LLM serving systems.

\section{BACKGROUND AND RELATED WORK}
\label{s:background}

\subsection{Multi-core AI Accelerators}

Recent AI accelerators increasingly adopt multi-core architectural designs, spanning single-chip, chiplet, and wafer-scale systems~\cite{IPU,NeuronCore,tenstorrent,DOJO,prabhakar2022sambanova,shao2019simba,firoozshahian2023mtia,lie2023cerebras,Groq}.

\begin{figure}[htp]
    \setlength{\abovecaptionskip}{0pt}
    \setlength{\belowcaptionskip}{-15pt}
    \includegraphics[width=\linewidth]{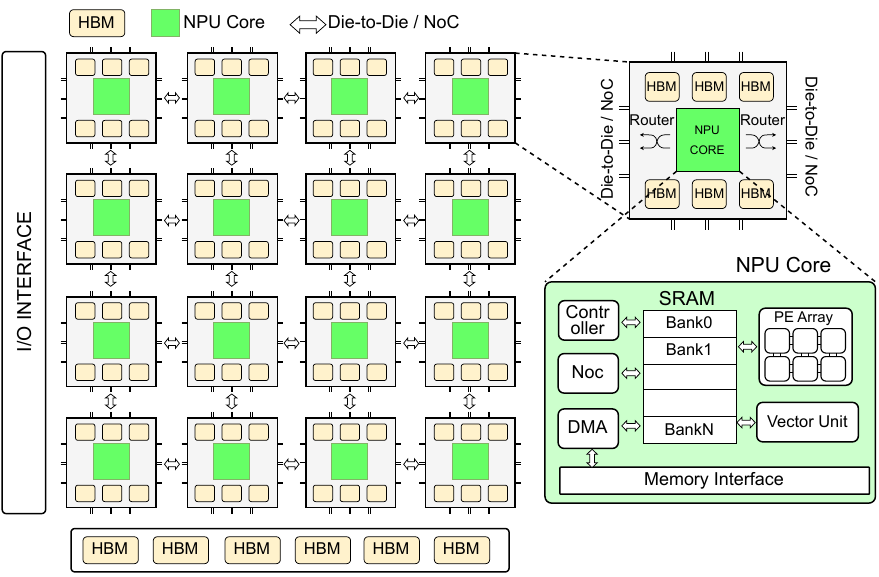}
    \caption{\textbf{Hardware architecture of multi-core AI accelerators.}}
    \label{fig:design-multi-core}
\end{figure}
The proliferation of these accelerators stems from DSE~\cite{10745857,10318183,10319731} that balances throughput, energy efficiency, and LLM memory-wall constraints, leading to diverse concrete configurations that still share a high-level organization.
As shown in Fig.~\ref{fig:design-multi-core}, a representative multi-core AI accelerator comprises compute cores, interconnect networks, on-chip/off-chip memory, and I/O interfaces, which provide computation, data movement, storage, and external communication support for LLM serving workloads.
Within each compute core, systolic arrays or matrix cubes, vector units, scalar units, local SRAM or scratchpad memory, DMA engines when external memory is present, and NoC routers are typically integrated.
The compute core is the smallest unit of computation scheduling, with multiple cores often integrated within a single chip or die.

As for the interconnect network, to balance hardware cost and scalability,
current multi-core AI accelerators frequently adopt a 2D-mesh topology that supports core-to-core, die-to-die, and chip-to-chip communication with high bandwidth and low latency across different scales.
However, this organization also makes tensor placement and communication patterns sensitive to physical core locations.

Memory subsystems of multi-core architectures vary substantially: early designs such as IPU~\cite{IPU} and Groq~\cite{Groq} typically use large SRAM as their sole memory resource, restricting single-chip support to small models; Cerebras WSE~\cite{lie2023cerebras} scales on-chip memory to the wafer level but still faces a mismatch between memory capacity and compute potential; and recent advances~\cite{prabhakar2022sambanova,wsc-llm} introduce external memory subsystems, including globally shared HBM or core-local HBM through memory stacking.
To accommodate LLM inference demands, multi-core AI accelerators will increasingly integrate high-speed, core-private memory resources, creating capacity, bandwidth, and data-movement trade-offs that motivate memory-aware LLM serving policies.

\newcommand{\cmark}{\textcolor{green!60!black}{\ding{51}}}
\newcommand{\xmark}{\textcolor{red}{\ding{55}}}
\newcommand{\pmark}[1]{\textcolor{orange}{\ding{109}}\, \textcolor{orange}{\small #1}}

\begin{table*}[]
	\setlength{\abovecaptionskip}{0pt}
	\caption{\textbf{Comparison of different simulators for AI accelerators:} This table compares various simulation tools based on their support for LLM serving (including PD disaggregation/fusion, and KV cache management),
	computational modeling (architectural vs. performance model), memory modeling (event vs. performance model), and capabilities for exploring new architectural design spaces (scale up network, on chip routing, hybrid bonding, dataflow).}
	\centering
	\resizebox{\textwidth}{!}{
        \setlength{\tabcolsep}{2pt}
		\label{tab:llm_sim_comparison}
		\begin{tabular}{@{}l|ccc|cc|cc|cccc@{}}
			\toprule
			\hline
			\multirow{2}{*}{\textbf{Simulator}} &
			\multicolumn{3}{c|}{\textbf{LLM Serving Support}} &
			\multicolumn{2}{c|}{\textbf{Computation Simulation}} &
			\multicolumn{2}{c|}{\textbf{Memory Simulation}} &
			\multicolumn{4}{c}{\textbf{New Architecture Exploration}} \\
			\cline{2-4} \cline{5-6} \cline{7-8} \cline{9-12}
			& P/D Disagg & P/D Fusion & KV Cache Mgmt & Micro-Arch & Perf/Profiler & TLM/Event & Perf/Profiler & Routing & Network & Hybrid-bonding & Dataflow \\ \hline
			\midrule
			RapidChiplet~\cite{iff2023rapidchiplet}    & \xmark & \xmark & \xmark & \xmark & \xmark & \xmark & \xmark & \cmark & \xmark & \cmark & \cmark \\ \hline
			h21lm~\cite{li2025h2}           		   & \xmark & \xmark & \xmark & \xmark & \cmark & \xmark & \cmark & \xmark & \xmark & \cmark & \xmark \\ \hline
			LLMServingSim~\cite{LLMServingSim}         & \cmark & \cmark & \cmark & \xmark & \pmark{Real-Machine} & \xmark & \pmark{Real-Machine} & \xmark & \xmark & \xmark & \cmark \\ \hline
			astra-sim~\cite{ASTRAsim2}                 & \xmark & \xmark & \xmark & \xmark & \cmark & \xmark & \cmark & \xmark & \cmark & \xmark & \cmark \\ \hline
			vidur~\cite{agrawal2024vidur}              & \cmark & \cmark & \pmark{Partial} & \xmark & \pmark{Real-Machine} & \xmark & \pmark{Real-Machine} & \xmark & \xmark & \xmark & \cmark \\ \hline
			apex~\cite{lin2024apex}                    & \xmark & \cmark & \xmark & \xmark & \pmark{Real-Machine} & \xmark & \pmark{Real-Machine} & \xmark & \xmark & \xmark & \cmark \\ \hline
			mNPUsim~\cite{mnpusim}                     & \xmark & \xmark & \xmark & \cmark & \cmark & \cmark & \cmark & \xmark & \xmark & \xmark & \xmark \\ \hline
			Timeloop~\cite{timeloop}                   & \xmark & \xmark & \xmark & \cmark & \cmark & \cmark & \cmark & \xmark & \xmark & \xmark & \xmark \\ \hline
			ONNXim~\cite{ONNXim}          			   & \xmark & \xmark & \xmark & \cmark & \cmark & \cmark & \cmark & \cmark & \xmark & \xmark & \xmark \\ \hline
			dcra~\cite{orenesvera2024dcra}             & \xmark & \xmark & \xmark & \xmark & \cmark & \xmark & \cmark & \cmark & \xmark & \cmark & \cmark \\ \hline
			\textbf{Ours}   & \cmark & \cmark & \cmark & \cmark & \cmark & \cmark & \cmark & \cmark & \cmark & \cmark & \cmark \\
			\bottomrule
		\end{tabular}
	}\\[-10pt]
\end{table*}

\newcommand{\green}{\color[HTML]{008000}}\begin{table*}[htp]
    \centering
    \setlength{\abovecaptionskip}{0pt}
    \setlength{\belowcaptionskip}{0pt}
    \caption{\textbf{Comparison of different methods for LLM inference.}}    \label{table:back:comparison}
    \centering\resizebox{\textwidth}{!}{
    \begin{tabular}{l|l|l|l|l|l|l|l}
        \toprule
        \hline
                            & \multicolumn{1}{c|}{\textbf{Tensor Partition}}                                                               & \multicolumn{1}{c|}{\textbf{Core Placement}}                                                        & \multicolumn{1}{c|}{\textbf{\begin{tabular}[c]{@{}c@{}}Memory \\ Management\end{tabular}}} & \multicolumn{1}{c|}{\textbf{\begin{tabular}[c]{@{}c@{}}Request \\ Scheduling\end{tabular}}} & \multicolumn{1}{c|}{\textbf{\begin{tabular}[c]{@{}c@{}}PD \\ Disaggregation\end{tabular}}} & \multicolumn{1}{c|}{\textbf{\begin{tabular}[c]{@{}c@{}}PD\\ Fusion\end{tabular}}} & \multicolumn{1}{c}{\textbf{Target Platform}} \\
        \midrule
        \textbf{T10}~\cite{T10}      & {\color[HTML]{CB0000} AllGather}                                                                             & {\color[HTML]{CB0000} Linear, 2D Mesh}                                                              & {\color[HTML]{CB0000} SRAM}                                                                & {\color[HTML]{CB0000} Not mentioned}                                                        & {\color[HTML]{CB0000} No}                                                                  & {\color[HTML]{CB0000} No}                                                         & {\color[HTML]{CB0000} IPU}                   \\
        \textbf{WaferLLM}~\cite{WaferLLM} & {\color[HTML]{F56B00} AllGather, AllReduce}                                                                  & {\color[HTML]{F56B00} \begin{tabular}[c]{@{}l@{}}Interleaved Linear, \\ 2D Mesh\end{tabular}}       & {\color[HTML]{F56B00} SRAM}                                                                & {\color[HTML]{CB0000} Not mentioned}                                                        & {\color[HTML]{CB0000} No}                                                                  & {\color[HTML]{CB0000} No}                                                         & {\color[HTML]{CB0000} Cerebras WSE}          \\
        \textbf{WSC-LLM}~\cite{wsc-llm}      & {\color[HTML]{CB0000} AllReduce}                                                                             & {\color[HTML]{CB0000} 2D Mesh}                                                                      & {\color[HTML]{F56B00} HBM}                                                                 & {\color[HTML]{008000} Yes}                                                                  & Fixed / DP-prioritized                                                             & {\color[HTML]{CB0000} No}                                                         & {\color[HTML]{F56B00} Wafer-scale Chip}      \\ \hline
        \textbf{Ours}      & {\color[HTML]{008000} \begin{tabular}[c]{@{}l@{}}AllGather, AllReduce, \\ AllGather+AllReduce\end{tabular}} & {\color[HTML]{008000} \begin{tabular}[c]{@{}l@{}}Interleaved Linear, \\ Ring, 2D Mesh\end{tabular}} & {\color[HTML]{008000} SRAM+HBM}                                                            & {\color[HTML]{008000} Yes}                                                                  & Simulation-guided / workload-aware                                                 & {\color[HTML]{008000} Yes}                                                        & {\color[HTML]{008000} Multi-core Chip}       \\ \hline
        \bottomrule
    \end{tabular}}\\[-5pt]
\end{table*}

\subsection{Simulators for AI accelerators}

As shown in Table~\ref{tab:llm_sim_comparison}, recent AI accelerator simulators mainly follow two methodologies: (1) cycle-accurate simulation and (2) performance simulation based on analytical models or real-hardware profiles, which largely trade off low-level fidelity against LLM-scale simulation efficiency.

\myparagraph{Cycle-accurate models} evaluate target chip architectures by simulating each clock cycle, commonly through cycle-by-cycle (cycle-loop) simulation or Register Transfer Level (RTL) simulation, as adopted by prior works~\cite{chen2016eyeriss,chen2014diannao,DaDianNao,EyerissV2,SCNN,Cnvlutin,MatRaptor,OuterSPACE,SIGMA,Cambricon,CambriconX,SpArch,STONNE,EIE}.
Although such modeling fully leverages low-level hardware details to obtain accurate hardware performance, its excessively long simulation time makes it ineffective for computationally demanding workloads such as LLMs or architectures with large-scale resources; FPGA acceleration~\cite{DNNBuilder,Gemmini,kim2023aurora} reduces this cost but remains limited by constrained FPGA resources, rendering end-to-end LLM serving simulation impractical.
\myparagraph{Performance models} use analytical equations or hardware profiles to estimate compute and memory latency~\cite{timeloop,ONNXim,ruby,sparseloop,samajdar2018scale,cimloop,samajdar2020systematic,mnpusim,AccelSim,heo2024neupims}, enabling faster design-space exploration, operator mapping, and architectural sweeps, but often abstracting away LLM-serving behavior; profile-driven variants are also hard to apply to emerging accelerators whose real systems are not yet available.
LLM serving differs fundamentally from traditional DNN inference because of its autoregressive computation pattern, 3D parallelism strategy, and PD disaggregation and fusion.
Recent works~\cite{LLMServingSim,ASTRAsim2,ASTRAsim,SimAI,agrawal2024vidur} attempt to simulate LLM serving features, but they typically rely on performance profiles from existing hardware and lack microarchitectural simulation capabilities for compute cores, limiting their ability to explore new hardware architectures.

In this paper, we present \sys, a framework that simultaneously supports multiple serving strategies for LLM deployment and enables the exploration of emerging hardware architectures, including NoC, dataflow architectures, hybrid bonding, 3D-stacked memory, etc.

\subsection{Accelerating LLM serving for Multi-core AI accelerators}

Prior research on multi-core AI accelerator architecture has predominantly optimized GEMM and GEMV computations, and Table~\ref{table:back:comparison} outlines the main strategies proposed in the literature.
T10~\cite{T10} optimizes matrix computation for the IPU~\cite{IPU} chip by introducing ``rotating tensors'', which distribute input and weight tensors across compute cores and use a rotating all-gather scheme~\cite{MLSYS2023_c4be71ab,SUMMA,cannon1969cellular} to collect complete matrix weights from other cores.
WaferLLM~\cite{WaferLLM} builds upon T10 and extends these techniques to the Cerebras WSE~\cite{lie2023cerebras} platform.
Because current multi-core AI accelerators typically employ a 2D-mesh topology, certain nodes may need to traverse up to $N$ hops to reach their logical neighbors during ring all-gather operations,
and WaferLLM addresses this with an interleaved arrangement that limits each ring all-gather to at most two hops.
These works show that tensor movement, topology-aware placement, and memory bandwidth are central to LLM serving on multi-core accelerators, but they still lack a comprehensive analysis of GEMM computation on multi-core architectures and remain platform-specific or omit parts of the tensor-partition, SRAM/NoC, and PD-fusion design space summarized in Table~\ref{table:back:comparison}.
To address these gaps, we conduct a systematic analysis of LLM serving acceleration techniques based on our dedicated simulator, including tensor partition, core placement, multi-level memory management, scheduling strategies, PD disaggregation, and PD fusion.

\section{\sys: a Multi-level Simulation Framework for Multi-core AI accelerators}
To better investigate the impact of different hardware configurations on LLM inference performance in multi-core AI accelerator architectures,
we have developed \sys, an efficient simulation platform for dataflow-based multi-core architectures.
\sys addresses two primary challenges: (1) efficiently and accurately simulating LLM inference tasks that are both computation-intensive and memory-intensive,
and (2) effectively handling streaming requests that are prevalent in contemporary LLM serving scenarios.

\begin{figure*}[htp]
    \centering
    \setlength{\abovecaptionskip}{-1pt}
    \setlength{\belowcaptionskip}{-10pt}
    \includegraphics[width=0.9\linewidth]{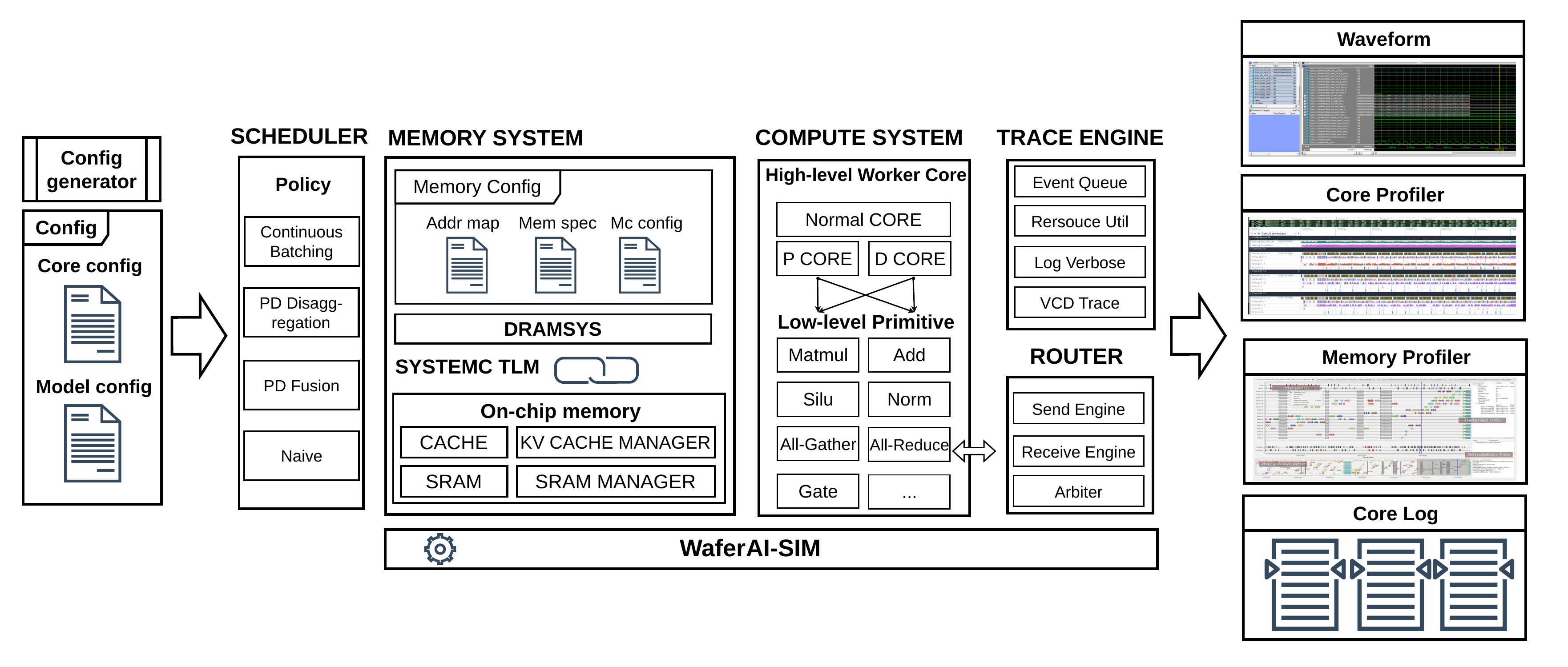}
    \caption{\textbf{The overall design of \sys:} Computing/memory/router sub-systems with tracing and scheduler models.}
    \label{fig:design:framework}
\end{figure*}
\subsection{Multi-level Simulation}
To balance simulation accuracy and speed, we employ a multi-level simulation approach as shown in Fig.~\ref{fig:design:framework}.
The entire simulation system is divided into three components: the computing system, the memory system, and the on-chip routing system.
Through careful analysis of these three components, we adopt a specific simulation level for each part.

For the computing system, we provide low-level primitive simulation implementations for various operators,
as well as high-level abstractions of worker cores (e.g., prefill cores).
Taking the Matmul operator as an example, we adopt a shape-aware performance model.
For a GEMM $(M_g \times K_g) \times (K_g \times N_g)$ executed on an $S \times S$ systolic array, we first partition the weights and input activations into tiles and pad the last tile if necessary.
The total computation latency is calculated as $T_{\text{comp}} = N_{\text{tiles}} \times T_{\text{cycles}} + T_{\text{inject}}$.

Here, $N_{\text{tiles}}$ is the number of Matmul tiles mapped to the systolic array,
$T_{\text{cycles}}$ is the per-tile systolic latency including reduction and fill/drain cycles,
and $T_{\text{inject}}$ is the one-time injection latency after the weight tile reaches the SRAM interface.
All terms are derived from architectural parameters,
following common simulators such as ONNXim~\cite{ONNXim} and ASTRA-sim~\cite{ASTRAsim2}.
Algorithm~\ref{alg:app-matmul} in the Appendix provides compact pseudocode for the Matmul compute and memory model.

For the memory system, prior works often employ empirical bandwidth-based equations to
estimate latency. However, high-bandwidth memory accesses exhibit characteristics such as
out-of-order execution, outstanding requests, and interleaving, so simple empirical equations fail to accurately
capture the true memory access latency.
To address this, we adopt a transaction-level modeling (TLM) approach~\cite{systemcTLM},
decomposing each memory request into four phases: \texttt{Begin\_Req}, \texttt{End\_Req}, \texttt{Begin\_Resp}, and \texttt{End\_Resp},
enabling asynchronous event-driven simulation.
This method achieves cycle-accurate simulation precision while maintaining high simulation efficiency.

For the routing system, arbitration, contention, deadlock-free guarantees, and routing decisions all affect on-chip throughput and data flow patterns, so we use cycle-accurate simulation with a handshaking mechanism to model router behavior.
Once a routing path is established through successful handshake exchange, one packet can be transmitted per clock cycle, allowing packet latency to be computed from the number of data transmissions over the established link.
Therefore, although the routing simulation is cycle-accurate, it does not significantly degrade the overall simulation speed.
\sys models communication cost from transferred payload, routed paths, and link contention,
and its routing model can be extended to non-mesh NoC topologies by specifying connectivity and routing policies.

\subsection{Customized Scheduler}
Previous works have primarily focused on CNN or static LLM simulations, in which a fixed batch of
requests is executed once to obtain the simulation runtime.
However, such simulation methodologies are significantly different from real-world LLM deployment scenarios.
For typical LLM scenarios, an end-to-end performance evaluation requires executing the prefill stage once, followed by the multiple decoding stages.
Simulation for LLM must handle dynamic graphs and scheduling,
where the sequence length during prefill, the number of decoding steps and the arrival time can vary across different requests.
We have implemented an iteration-level scheduler and monitor, which allows flexible configuration,
such as the number of requests per iteration, prompt length, chunked prefill, and prefill-decoding stages.
This design enables customized scheduling strategies, such as PD fusion, PD disaggregation, and continuous batching.
Details are provided in \textsection\ref{subsec:pd_disaggregation_fusion}.
The formal PD fusion and PD disaggregation scheduler pseudocode is provided in Algorithm~\ref{alg:app-pd-schedulers} in the Appendix.

\section{Optimizing LLM Serving Systems on Multi-core AI Accelerators using \sys}
\subsection{On-chip Distributed Systems}

Although multi-core AI accelerators such as the Cerebras WSE and Graphcore IPU have achieved SOTA decoding performance for LLM inference,
they rely solely on SRAM as their memory resource,
which necessitates cascading multiple chips to run a single model.
This design results in a substantial mismatch between compute power and memory capacity, leading to extremely low computing resource utilization.
To mitigate this issue, contemporary multi-core AI architectures tend to adopt 2.5D/3D memory stacking, hybrid bonding,
to provide additional memory media (e.g., HBM).
However, how to fully exploit both the computing (multi-core) and memory (hierarchical-memory) resources simultaneously remains a major challenge.
In our research, we conduct extensive explorations based on \sys to address these complexities, which include:

\begin{myitemize}
    \item \textbf{Drawbacks of multi-core architecture:} As the compute-core count continues to scale with distributed-memory architectures,
    the distributed tensor computation becomes the key factor for the LLM serving performance.
    \textsection\ref{sub:design:tp} proposes different tensor partitioning and placement strategies to accommodate various task scenarios.
    \item \textbf{Complexity of hierarchical memory system:} As models grow, a single chip's on-chip SRAM cannot store all the weights,
    necessitating the offloading of weights and the KV caches to off-chip memory, such as HBM.
    \textsection\ref{sub:design:mem} proposes a fine-grained, multi-level memory-object management system that fully exploits the high bandwidth of on-chip SRAM
    and the high capacity of off-chip HBM.
    \item \textbf{Challenge of heterogeneity:}
    Multi-core architectures naturally support heterogeneous core designs;
    however, contemporary multi-core AI accelerators predominantly employ homogeneous cores.
    \textsection\ref{subsec:pd_disaggregation_fusion} investigates how to leverage the heterogeneous characteristics of multi-core architectures
    to design efficient PD strategies that enhance the performance of LLM serving.
\end{myitemize}

\subsection{Tensor Management on Multi-core Accelerators}
Due to the dataflow computing paradigm of multi-core AI accelerator chips,
their performance is highly sensitive to tensor partition and placement strategies.
Although prior work has studied multi-core GEMM deployment, real-world LLM serving involves diverse model sizes and sequence lengths, making one-size-fits-all decisions inadequate.
\sys places these optimization choices in a unified simulation environment to evaluate their effectiveness across different hardware configurations, interconnect organizations, and serving workloads.

\myparagraph{Different tensor partition strategies:}
\label{sub:design:tp}
Unlike tensor-parallelism in GPUs,
the GEMM operation in multi-core architectures faces SRAM-capacity constraints that require partitioning both activation and weight tensors.

\begin{figure}[htp]
    \setlength{\abovecaptionskip}{-1pt}
    \setlength{\belowcaptionskip}{-15pt}
    \includegraphics[width=\linewidth]{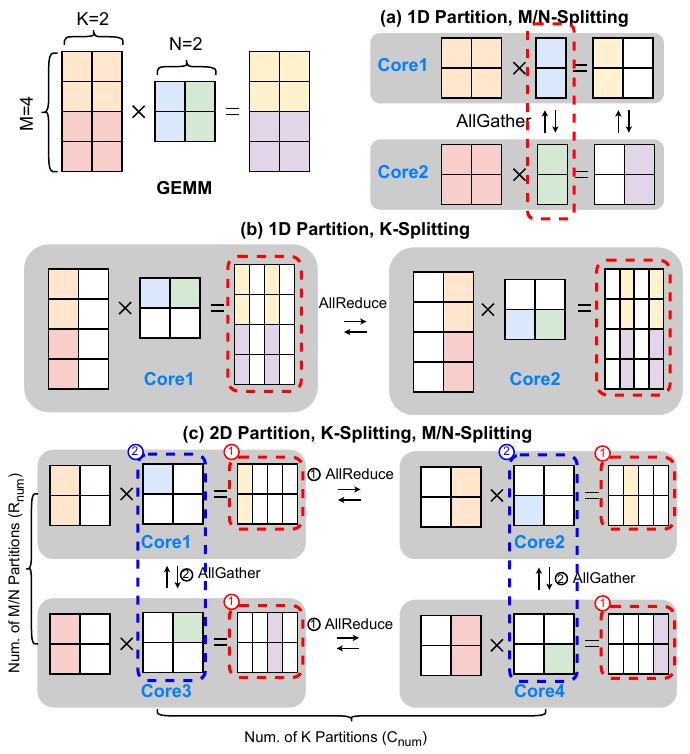}
    \caption{\textbf{Different tensor partition strategies:} For the GEMM operator, we consider three partition strategies:
    (1) 1-D partition along the M and N dimensions, which relies on the AllGather primitive;
    (2) 1-D partition along the K dimension, based on the AllReduce primitive;
    and (3) 2-D partition across the M/N and K dimensions, which leverages both AllReduce and AllGather primitives.}
    \label{fig:design-tp-partition}
\end{figure}

Tensor partitioning involves two aspects: (1) which dimensions and (2) how many dimensions to partition.
If activations and weights are respectively partitioned along the M and N axes,
each compute core must employ the AllGather primitive to gather the complete weight tensor from other cores.
By contrast, partitioning along the K dimension uses the AllReduce primitive to aggregate the partial results
(unlike T10's AllGather-style rotating-tensor execution~\cite[Sec.~4.2]{T10}, this reduction-based formulation is consistent with WaferLLM's ReduceAdd design for K-dimension distributed GEMM~\cite[Sec.~5.4]{WaferLLM}).
Fig.~\ref{fig:design-tp-partition}-a and Fig.~\ref{fig:design-tp-partition}-b illustrate these two partition approaches.

For 1-D tensor partition, GEMM cannot complete in a single iteration;
we use conventional ring AllGather or AllReduce to collect the complete tensor or result within $N$ iterations, where $N$ equals the number of partitions,
as shown in Fig.~\ref{fig:design-tp-partition}-a and Fig.~\ref{fig:design-tp-partition}-b.
For 2-D partition, tensors are partitioned along both M/N and K dimensions, and compute cores form a 2D mesh as shown in Fig.~\ref{fig:design-tp-partition}-c.
Each core repeatedly performs row-wise AllReduce for partial-result aggregation and column-wise AllGather for exchanging partial input or weight tensors until the computation completes.

Table~\ref{table:design:partition} exhibits the theoretical communication overhead,
maximum hops and memory cost of different tensor partition strategies mentioned above.

\begin{table*}
    \centering
    \setlength{\belowcaptionskip}{0pt}
    \setlength{\abovecaptionskip}{0pt}
    \caption{\textbf{Communication and memory cost of different tensor partition strategies.}
    \emph{Input/Weight/Output Tensor}, \emph{Total Communication}, and \emph{Max Hop} denote per-core memory cost, total inter-core transferred data during GEMM, and maximum transfer hops, respectively;
    $Num$, $R_{num}$, and $C_{num}$ denote the numbers of total, row, and column partitions.}
    \label{table:design:partition}
    \begin{tblr}{
        width = \textwidth,        hline{1,6} = {-}{0.08em},
        hline{2} = {-}{},
        colspec = {X[1.6, l] X[0.9, c] X[1, c] X[1,c] X[4.7, c] X[0.7, c]},    }
                        & \textbf{Input Tensor}     & \textbf{Weight Tensor}     & \textbf{Output Tensor}     & \textbf{Total Communication}                                                                                 & \textbf{Max Hop} \\
    Input-only Partition & $\frac{input\_size}{num}$ & $weight\_size$                & $\frac{output\_size}{num}$ & 0                                                                                                   & 0                         \\
    1-D Partition (M/N)  & $\frac{input\_size}{num}$ & $\frac{weight\_size}{num}$ & $\frac{output\_size}{num}$ & $\frac{num-1}{num}\times\begin{pmatrix}K \times N\end{pmatrix}$                                                 & $1 \sim 2$                        \\
    1-D Partition (K)    & $\frac{input\_size}{num}$ & $\frac{weight\_size}{num}$ & $\frac{output\_size}{num}$ & $2\times\frac{num-1}{num}\times\begin{pmatrix}M\times N\end{pmatrix}$                                               & $1 \sim 2$                        \\
    2-D Partition       & $\frac{input\_size}{R_{num}\times C_{num}}$ & $\frac{weight\_size}{R_{num}\times C_{num}}$ & $\frac{output\_size}{R_{num}\times C_{num}}$ & $(R_{num}-1)\times(2\times\frac{C_{num}-1}{C_{num}}\times\frac{M\times N}{C_{num}\times C_{num}}+\frac{K\times N}{C_{num}\times R_{num}})$ & $1 \sim 2$
    \end{tblr}\\[-15pt]
\end{table*}

\myparagraph{Different tensor placement strategies:}
Tensor placement is also crucial for distributed tensor computation.
We divide compute cores into multiple pipelines, each processing one or more model layers, and apply different intra-pipeline placement strategies for 1-D or 2-D tensor partitioning, including ring and sequence layouts.

\begin{figure}[htp]
    \setlength{\abovecaptionskip}{0pt}
    \setlength{\belowcaptionskip}{-10pt}
    \includegraphics[width=\linewidth]{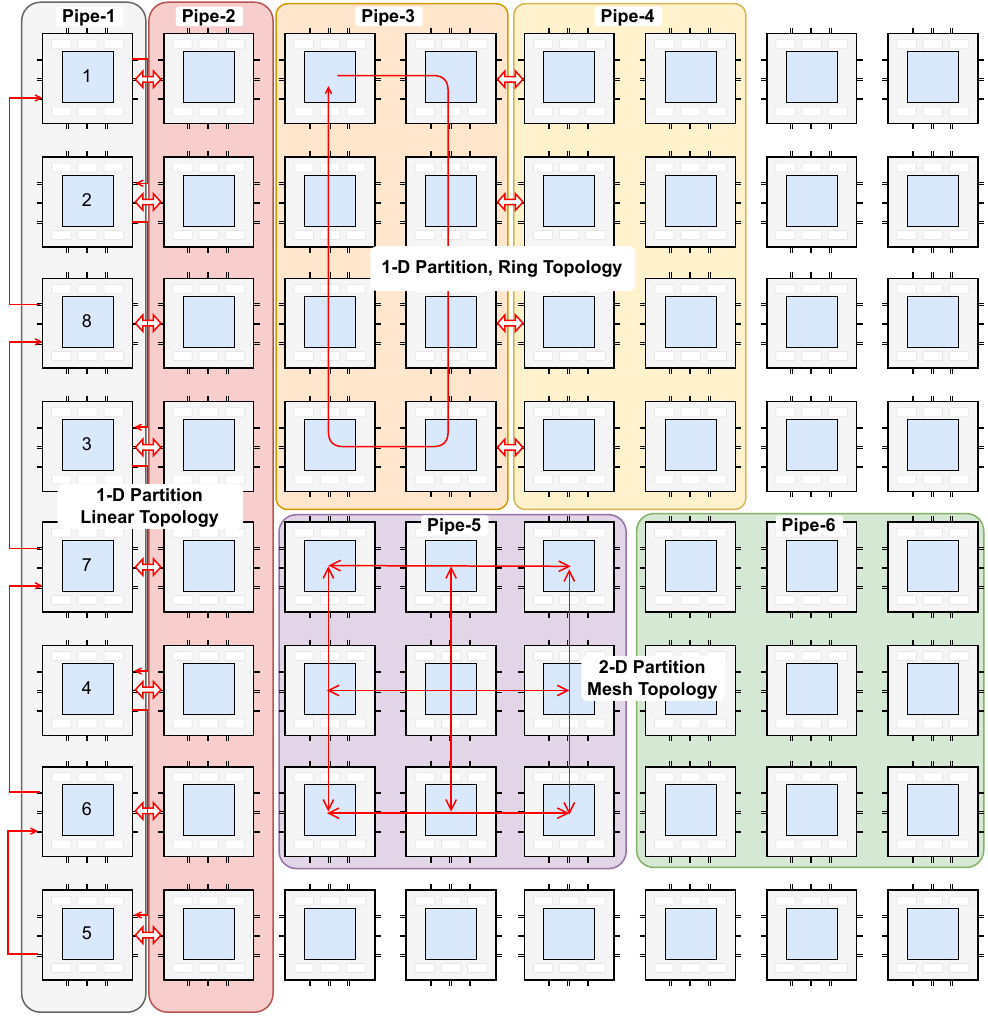}
    \caption{\textbf{Different tensor placement strategies:} Considering both inter-pipe and intra-pipe communication cost.}
    \label{fig:design-core-placement}
\end{figure}

Under 1-D placement/partition, each TP group can use either a ring topology, such as Pipe-3/4 in Fig.~\ref{fig:design-core-placement}, or WaferLLM's interleaved linear topology~\cite{WaferLLM}, such as Pipe-1/2.
The ring topology matches ring-based AllGather and AllReduce but reduces inter-pipeline bandwidth, whereas the interleaved linear topology improves inter-pipeline bandwidth at the cost of longer physical distances between logically adjacent virtual-ring nodes.
For 2-D placement, Pipe-5/6 organizes cores as a 2D mesh to increase intra-TP-group interconnections.

\subsection{Hierarchical Memory Management}
\label{sub:design:mem}
Recent multi-core accelerators combine on-chip SRAM with external HBM, creating hierarchical-memory challenges for KV-cache management and weight/activation placement during LLM serving.
To exploit SRAM performance and HBM capacity, we propose a hybrid-granularity memory management system that orchestrates memory objects across the hierarchy.

\begin{figure}[htp]
    \setlength{\abovecaptionskip}{-1pt}
    \setlength{\belowcaptionskip}{-15pt}
    \includegraphics[width=\linewidth]{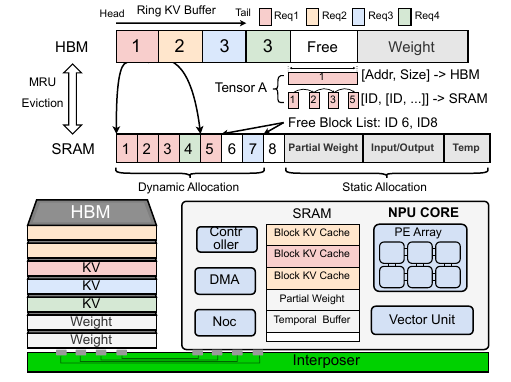}
    \caption{Multi-grained KV cache management for different memory hierarchies in AI accelerators. }
    \label{fig:design-kv-cache}
\end{figure}

\myparagraph{KV cache management:}
LLM serving memory objects include KV caches, model weights, inputs/activations, and temporary buffers.
KV caches require dynamic allocation because each request's generated length is unknown at initialization.
Due to limited SRAM, KV-cache blocks are managed at fine granularity on chip, while overflowed blocks spill to an HBM coarse-grained ring buffer for higher-capacity sequential access.
As shown in Fig.~\ref{fig:design-kv-cache}, request 1 grows block by block, and requests 2 and 3 later create interleaved, non-contiguous allocations.
We track each request's KV tensor with a block-ID linked list and recycle completed requests' blocks through a free-block list.

\myparagraph{Weight and activation management:}
During prefill, GEMM dominates performance, so keeping entire model weights/tensors in SRAM brings limited benefit; we therefore reserve only partial tensors in SRAM.
In contrast, inter-core communication data such as activations must reside entirely in SRAM.
We adopt MRU eviction because LLM inference traverses weights cyclically layer by layer, making recently consumed blocks less likely to be reused immediately.
When activation, weight, and KV-cache demand approach their limits simultaneously, the pressure is reflected as additional HBM traffic or infeasible configurations.

\subsection{Heterogenous Management for PD Stages}
\label{subsec:pd_disaggregation_fusion}
During LLM serving on multi-core AI accelerators, resource utilization is imbalanced between prefill and decoding, making PD disaggregation or PD fusion effective.
The heterogeneity-friendly nature of multi-core architectures further opens opportunities for improving serving performance.

\begin{figure}[htp]
    \setlength{\abovecaptionskip}{-1pt}
    \setlength{\belowcaptionskip}{-15pt}
    \includegraphics[width=\linewidth]{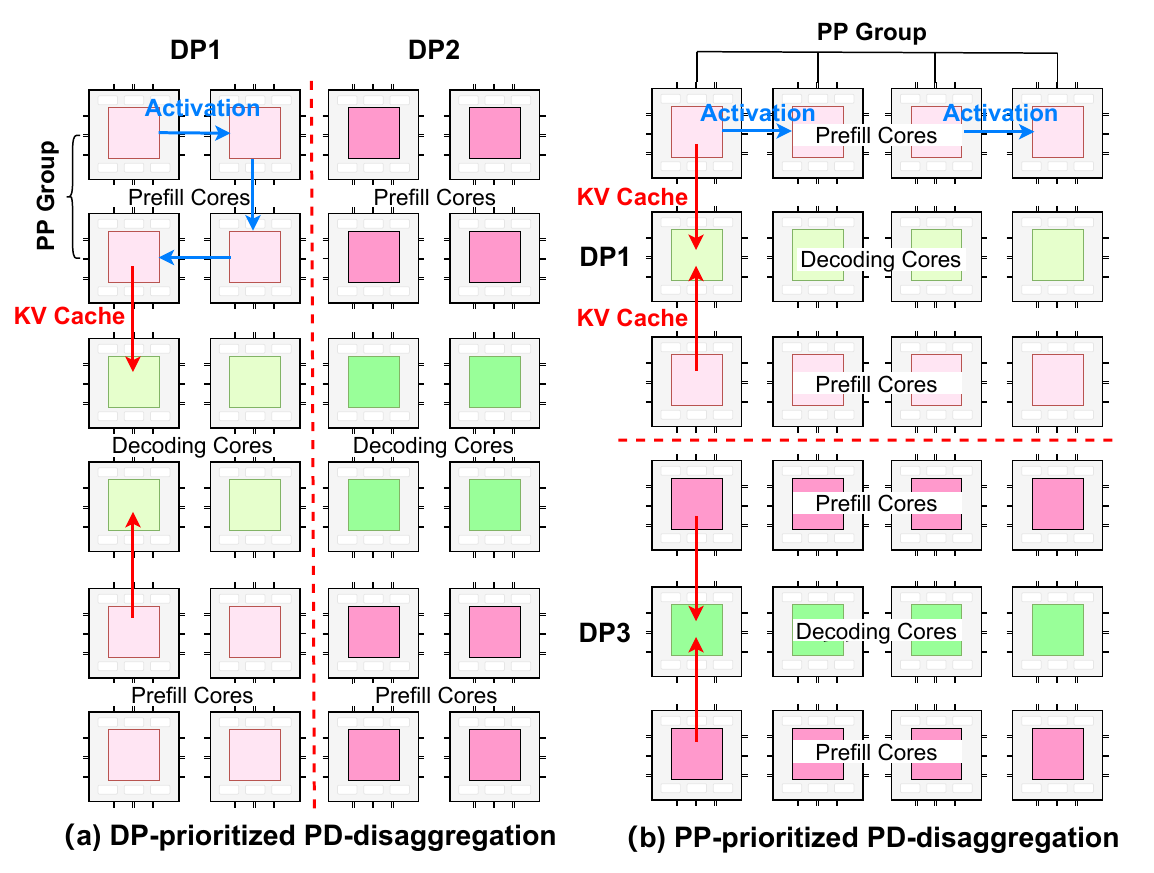}
    \caption{\textbf{Different PD disaggregation strategies:} Figure (a) illustrates the DP-prioritized core placement strategy for PD disaggregation;
    Figure (b) illustrates the PP-prioritized core placement strategy.}
    \label{fig:design-pd-disaggregation}
\end{figure}

We first consider homogeneous-core PD disaggregation.

\myparagraph{Core placement for PD disaggregation:}
For PD disaggregation, on-chip cores can be flexibly split between prefill and decoding.
Prior work~\cite{wsc-llm} uses DP-prioritized placement, where cores are first grouped by data parallelism and then assigned to prefill or decoding according to a specified ratio, as shown in Fig.~\ref{fig:design-pd-disaggregation}-(a).
We instead prioritize pipeline parallelism: because pipeline-parallel execution uses only one interconnect channel in the 2D mesh, the remaining channels can transfer KV caches from prefill cores to decoding cores.
As shown in Fig.~\ref{fig:design-pd-disaggregation}-(b), this PP-prioritized placement increases prefill-to-decoding bandwidth without affecting pipeline-parallel data transfer.
For a fixed workload, DP-prioritized and PP-prioritized placements transfer the same KV-cache payload but differ in routing distance and link contention.

\myparagraph{Heterogeneous parallel strategies for PD disaggregation:}
PD disaggregation also requires selecting parallel strategies for prefill and decoding, such as TP and PP sizes.
Prefill requests can stream into prefill cores without waiting for previous tasks, enabling efficient pipeline parallelism.
Decoding is autoregressive, so pipeline parallelism increases both decoding latency and per-core KV-cache reservation by \textbf{N-fold}, where N is the number of pipeline stages.
Tensor parallelism is therefore preferable for decoding because it aggregates memory bandwidth across cores and reduces per-core KV-cache allocation.

\myparagraph{Heterogeneous core design for PD disaggregation:}
\sys studies heterogeneous PD designs together with serving policies, exposing feasibility, configuration space, and fusion/disaggregation trade-offs before hardware commitment.
Because prefill and decoding have distinct computational characteristics, assigning heterogeneous cores to each stage can be beneficial.

Reallocating decoding-core resources from compute to memory has limited impact on decoding GEMM due to small batch sizes, while improving GEMV performance and supporting more decoding requests.

\myparagraph{PD Fusion on multi-core AI accelerators:}
In \sys, PD disaggregation and PD fusion are scenario-level serving configurations, with schedulers managing request-state transitions.
For PD fusion, we use a scheduler that co-locates prefill and decoding while satisfying TBT (Time Between Token) and TTFT (Time To First Token) requirements.
To limit prefill interference with decoding, we adopt chunked prefill~\cite{agrawal2023sarathiefficientllminference}: each core has a fixed budget, each decoding task consumes one unit, and each prefill chunk consumes $N$ units.
The scheduler prioritizes decoding when decoding demand exceeds the budget, and otherwise assigns available slots to chunked prefill.

\section{EVALUATION}
\label{s:eval}

\subsection{Experiment Setup}
We first validate the accuracy and efficiency of \sys.
\sys integrates modules from existing simulators, such as ONNXim~\cite{ONNXim} and Dramsys~\cite{steiner2020dramsys4}.
Based on \sys, we test different serving strategies on various LLM models and workloads.
Meanwhile, we also present guidelines for the architectural design of multi-core AI accelerators based on our results.

\begin{table}[t]
  \centering
  \small
  \setlength{\tabcolsep}{6pt}
  \setlength{\abovecaptionskip}{0pt}
  \setlength{\belowcaptionskip}{0pt}
  \caption{Chip configuration space for evaluation.}
  \label{tab:chip-config-space}
  \resizebox{\columnwidth}{!}{\begin{tabular}{|l|c|c|}
    \toprule
    \textbf{Parameter} & \textbf{Large-core} & \textbf{Small-core} \\
    \midrule
    \# of compute cores & 64 & 256 \\
    Core frequency & 500\,MHz & 500\,MHz \\
    Systolic array size & $32{\times}32$--$128{\times}128$ & $32{\times}32$--$64{\times}64$ \\
    Vector unit (64 ALUs/lane) & 32--128 lanes & 32--64 lanes \\
    SRAM per core & 8--128\,MB & 8--48\,MB \\
    SRAM bandwidth per core & scaled with SA & scaled with SA \\
    NoC bandwidth & 16--480\,GB/s${\times}$4  & 8--160\,GB/s${\times}$4 \\
    HBM bandwidth per core & 30--480\,GB/s & 15--60\,GB/s \\
    \bottomrule
  \end{tabular}
}\\[-10pt]
\end{table}

\myparagraph{Chip configurations:}
We consider a variety of hardware configurations for multi-core AI accelerators, as summarized in Table~\ref{tab:chip-config-space}.
These configurations encompass compute-core count, compute capability, SRAM size and bandwidth, HBM capacity and bandwidth, among other parameters.

\myparagraph{Model selection:}
We evaluate dense Qwen3 and Llama-3 models from 1.7B to 32B and MoE models from 30B to 235B for completeness.
For MoE, we follow expert-parallel load balancing (EPLB)~\cite{eplb}: experts are placed by estimated routing load, frequently activated experts are replicated to reduce hot spots, and gating/router operators assign tokens to experts. \sys models the resulting NoC/distributed-memory dispatch/combine traffic, including extra communication from expert replication and load-balanced routing.

\myparagraph{Workloads:}
In LLM serving scenarios, we follow industrial traces such as~\cite{Sharegpt-trace} and~\cite{mooncake-trace}.
Guided by these traces, we adopt three representative workloads: \textbf{TriviaQA}~\cite{TriviaQA}, with an average of 1,787 input tokens and 5 output tokens;
\textbf{GSM8K}~\cite{GSM8K}, with an average of 296 input tokens and 340 output tokens;
and \textbf{MultiTurn Chat}~\cite{MultiturnChat}, with an average of 54 input tokens and 374 output tokens. These workloads respectively characterize prefill-dominated, balanced prefill-decode, and decode-dominated serving patterns.

\begin{figure*}[ht]
    \centering
    \setlength{\abovecaptionskip}{-1pt}
    \setlength{\belowcaptionskip}{-18pt}
    \includegraphics[width=\linewidth]{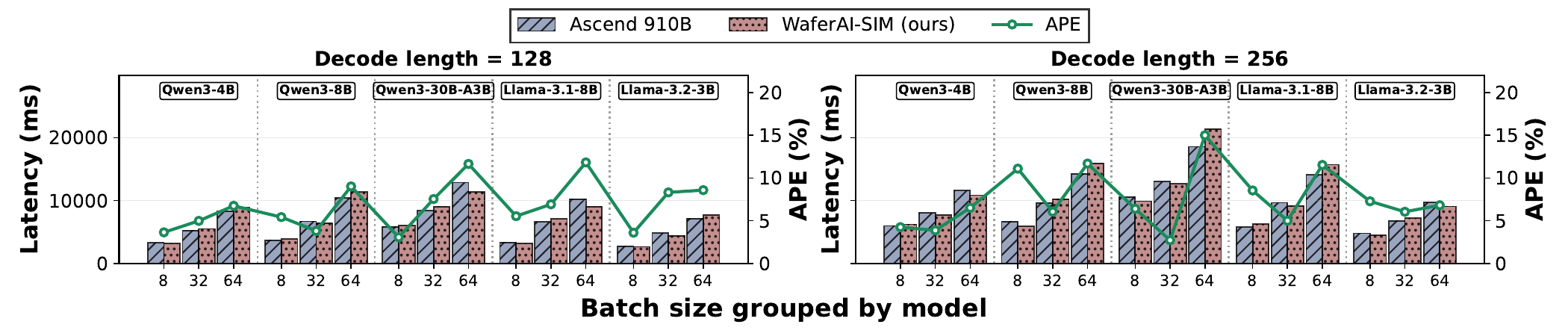}
    \caption{End-to-end latency and simulation error of \sys against Ascend-NPU-910B hardware.
    Bars report Ascend-NPU-910B and \sys latency in milliseconds, while the line reports the absolute percentage error (APE) relative to hardware.
    Subplots (a) and (b) show decode sizes of 128 and 256, respectively.}
  \label{fig:simulation_perf}
\end{figure*}

\begin{figure}[ht]
    \centering
    \setlength{\abovecaptionskip}{-1pt}
    \setlength{\belowcaptionskip}{-15pt}
    \includegraphics[width=\linewidth]{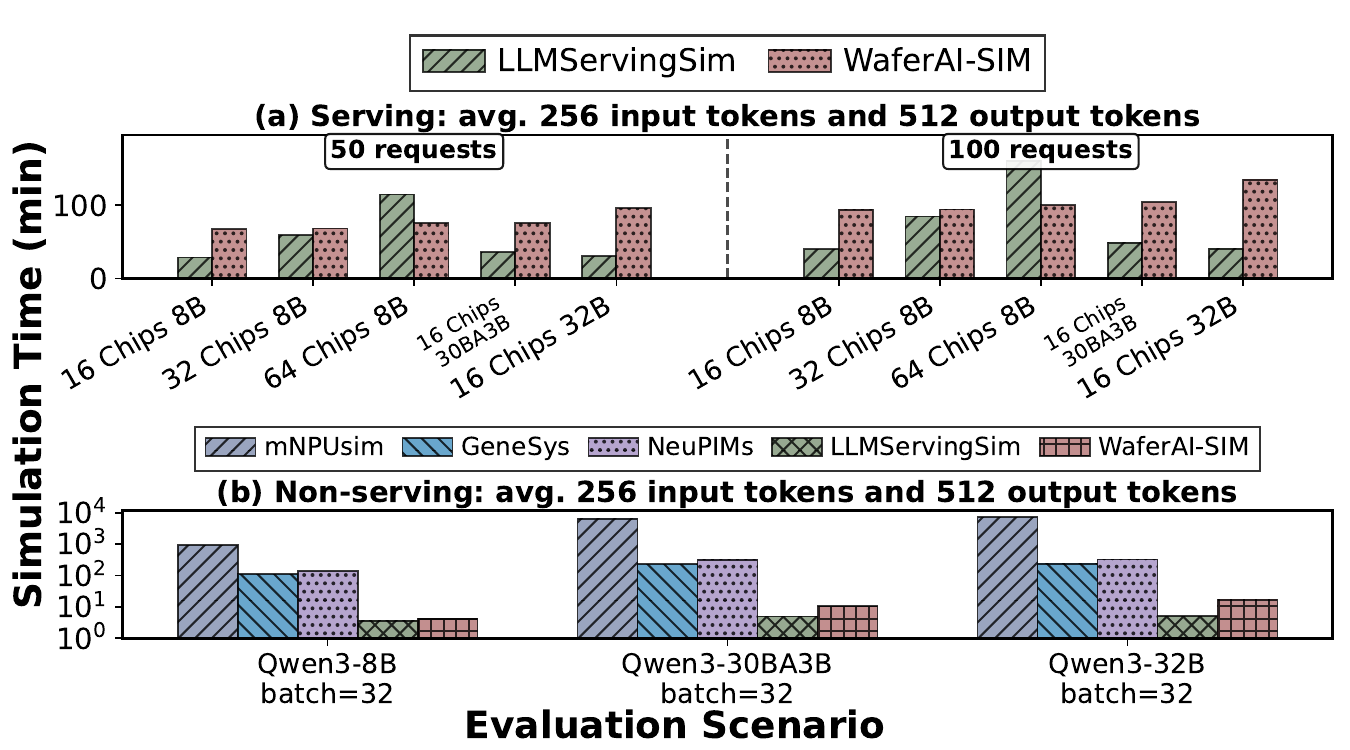}
    \caption{Wall-clock simulation time compared with other simulators under matched serving and non-serving scenarios. \textbf{Notably, LLMServingSim further requires hours of additional real-hardware profiling time before simulation.}}
    \label{fig:serving_wallclock}
\end{figure}

\begin{table}[tp]
\centering
\caption{Simulation accuracy of \sys relative to Ascend-NPU-910B hardware.
MAPE and maximum error are computed across all tested batch sizes and decode sizes for each model.}
\label{tab:simulation_accuracy}
\resizebox{0.75\linewidth}{!}{
\begin{tabular}{lcc}
\toprule
Model & MAPE (\%) & Max error (\%) \\
\midrule
Qwen3\_4B & 4.48 & 6.70 \\
Qwen3\_8B & 5.43 & 8.10 \\
Qwen3\_30B-A3B & 7.15 & 10.40 \\
Llama-3.1-8B & 5.52 & 8.40 \\
Llama-3.2-3B & 3.88 & 5.60 \\
\midrule
Overall & 5.29 & 10.40 \\
\bottomrule
\end{tabular}}
\end{table}

\subsection{Simulator Fidelity Across Latency and Runtime}
Fig.~\ref{fig:simulation_perf} and Fig.~\ref{fig:serving_wallclock} evaluate \sys in latency fidelity and simulation cost.
Against the real Ascend-NPU-910B hardware~\cite{DaVinci}, \sys tracks end-to-end latency trends with an overall MAPE of 5.29\% and a maximum error of 10.40\% (Table~\ref{tab:simulation_accuracy}); residual errors mainly stem from hardware-specific operators, runtime scheduling, and opaque software optimizations.
Compared with the serving-oriented simulator LLMServingSim~\cite{LLMServingSim}, \sys is slower at small emulated compute-core counts but scales more gently, becoming faster at the largest compute-core counts; in the non-serving single-batch case, \sys also remains competitive against multiple other simulators.

\myparagraph{Threat to validity and internal cross-check:}
We validate on Ascend-NPU-910B because its driver, hardware configuration, and related architecture details are public; validating other accelerators requires concrete hardware parameters and software behavior, including memory-hierarchy access, NoC routing/interconnect, runtime scheduling, kernels, driver-level memory behavior, and tensor-core-to-memory data movement.
We cross-check the results with Ascend-NPU-910B end-to-end simulation and a hardware-calibration-independent NoC communication-volume analysis: under K-dimension partitioning on $P$ cores, the analytical AllGather-over-ring-AllReduce volume ratio is $P/2$. In Fig.~\ref{fig:tp_kdim_comm}, the simulated latency ratio follows this trend, rising from $1.29\times$ at TP=4 to $6.15\times$ at TP=16, but remains below the pure volume ratio because the theory counts only payload volume and omits compute/memory interactions and dynamic NoC effects such as congestion and deadlock-prevention overhead.

\begin{figure*}[ht]
    \centering
    \setlength{\abovecaptionskip}{-1pt}
    \setlength{\belowcaptionskip}{-15pt}
    \includegraphics[width=\textwidth]{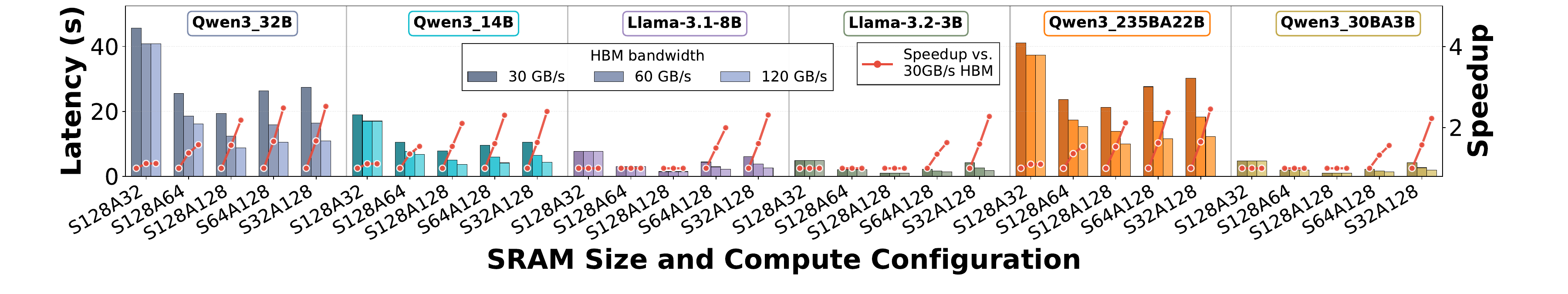}
    \caption{\textbf{Single-request latency of Qwen3 and Llama-3 models under varying hardware configurations.}
    All experiments use a 64-compute-core configuration with TP=4 and the prefill-decoding token ratio of 5:1.
    The corresponding DP and PP configurations are automatically determined to maximize resource utilization.
    The x-axis indicates SRAM-compute configurations (e.g., ``S32A12'' denotes 32\,MB SRAM and a $128 \times 128$ systolic array).}
    \label{fig:qwen_bundle}
\end{figure*}

\subsection{Hardware Configuration Space Exploration}
Fig.~\ref{fig:qwen_bundle} presents single-request latency for Qwen3 and Llama-3 models under varying hardware configurations,
including the per-core SRAM size, systolic array dimension, and HBM bandwidth.

In prefill-dominated workloads, scaling compute resources can significantly reduce latency.
Using Qwen-32B as an example, expanding the systolic array from $32\times32$ to $64\times64$ decreases end-to-end latency by approximately 55\%.
A subsequent expansion to $128\times128$ yields a smaller latency improvement of about 33\%.
Although larger compute cores can deliver higher performance, they may also lead to wasted compute resources, for example due to input padding.

For smaller models (e.g., 3B--8B), larger SRAM keeps nearly all weights and KV cache on-chip, reducing off-chip accesses and latency.
However, cloud-scale deployments use much larger models, making sufficient on-core SRAM prohibitively expensive.
Frequent off-chip spilling shifts the bottleneck to HBM bandwidth: increasing per-core bandwidth from 30 GB/s to 120 GB/s yields up to a 2.5X speedup for a 32B model, whereas additional SRAM offers limited benefit.

In summary, large SRAM is required only when users pursue extremely high decoding performance
since this approach needs to store all model weights in SRAM.
In other scenarios, the size of SRAM is not the primary determinant of LLM serving performance.

\begin{figure}[ht]
    \setlength{\abovecaptionskip}{0pt}
    \setlength{\belowcaptionskip}{-12pt}
    \includegraphics[width=\linewidth]{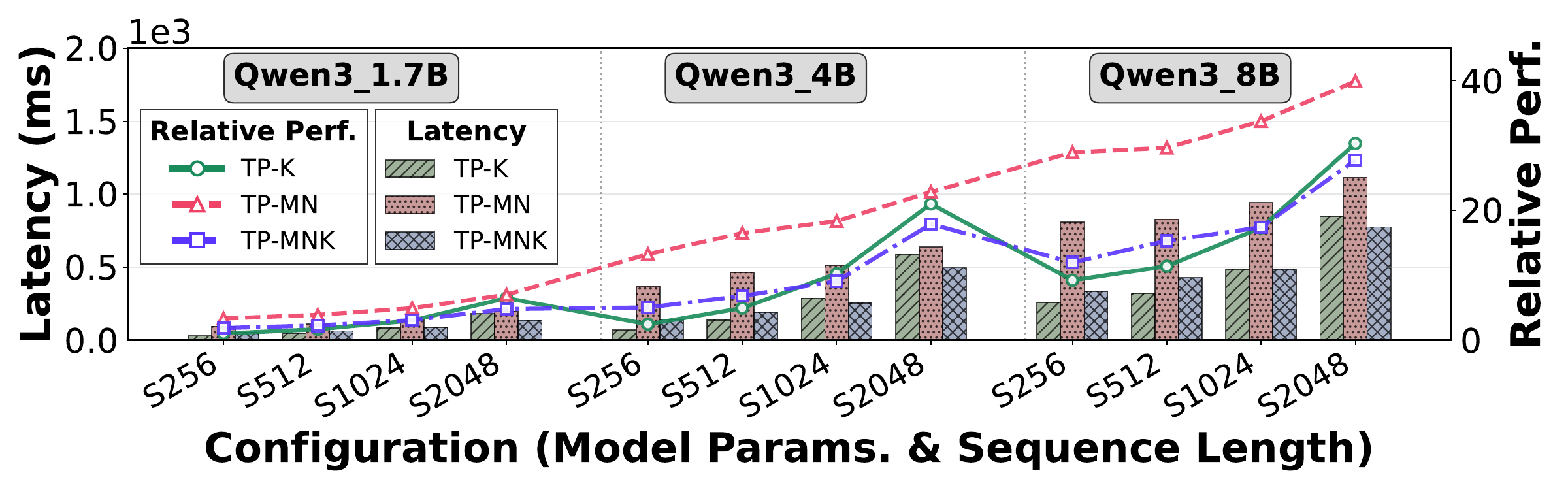}
    \caption{Impact of TP partition strategies on request latency across varying input sequence lengths.}
    \label{fig:tp_dimension}
\end{figure}

\subsection{TP and Core Placement}

\myparagraph{Different TP partition strategies:}
Fig.~\ref{fig:tp_dimension} compares TP=4 latency as input length varies across TP partition strategies.
K-dimension partition performs best when the input length is below the model hidden dimension; for Qwen3\_4B at sequence length 256, it is 6.03x faster than MN-dimension partition.
In long-context cases, once the sequence length exceeds the hidden dimension, K-dimension performance degrades sharply.
MNK 2D partitioning outperforms MN 1D partitioning by 1.44x on average, consistent with our theoretical analysis(\textsection\ref{sub:design:tp}).
Figure~\ref{fig:tp_kdim_comm} in the Appendix isolates AllGather versus AllReduce under K-dimension partitioning while fixing all other variables.

\begin{figure}[htp]
    \setlength{\abovecaptionskip}{0pt}
    \setlength{\belowcaptionskip}{-15pt}
    \includegraphics[width=\linewidth]{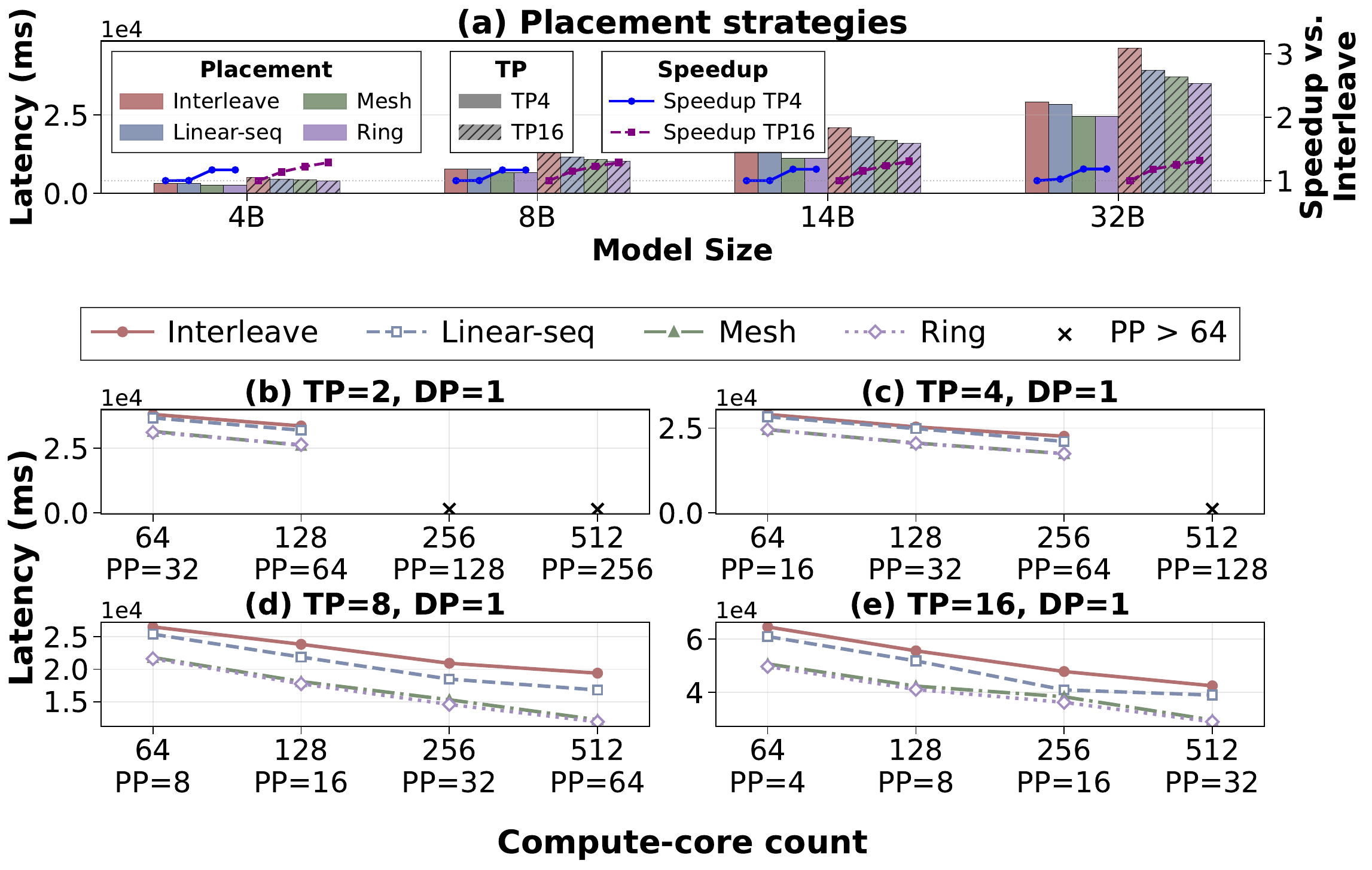}
    \caption{Single-request latency under different core placements.
Subplot (a) compares model sizes at TP=4/16; subplots (b)--(e) evaluate Qwen3-32B at TP=2/4/8/16 while scaling the total compute-core count from 64 to 512 with DP=1.
Configurations requiring PP$>$64 are infeasible for the 32B model and marked by ``x''.}
    \label{fig:tp_mapping}
\end{figure}

\myparagraph{Core placement strategy:}
Fig.~\ref{fig:tp_mapping} reports single-request latency under different placements: linear-seq follows the T10~\cite{T10} core-index order, while linear-interleave follows the Wafer-LLM~\cite{WaferLLM} mapping that limits each transmission to at most two hops.
For TP=4, linear-interleave and linear-seq are comparable, whereas mesh and ring topologies achieve a 1.17x speedup.
At TP=16, placement optimization becomes more important: relative to linear-interleave, linear-seq, and mesh, ring yields maximum speedups of 1.18x, 1.25x, and 1.32x, respectively.
As TP increases, performance first improves and then degrades because higher intra-layer parallelism is eventually offset by larger-TP communication overhead.
Although Wafer-LLM reports linear-interleave as best on Cerebras, its effectiveness may differ on other hardware; in our implementation, deadlock-free channel locking reduces interleaved-communication performance, making mesh and ring mappings more effective.
The placement-strategy ranking remains stable as the compute-core count scales up.

\subsection{SRAM Size and Memory Management}
\begin{figure}[ht]
    \setlength{\abovecaptionskip}{0pt}
    \setlength{\belowcaptionskip}{-10pt}
    \includegraphics[width=\linewidth]{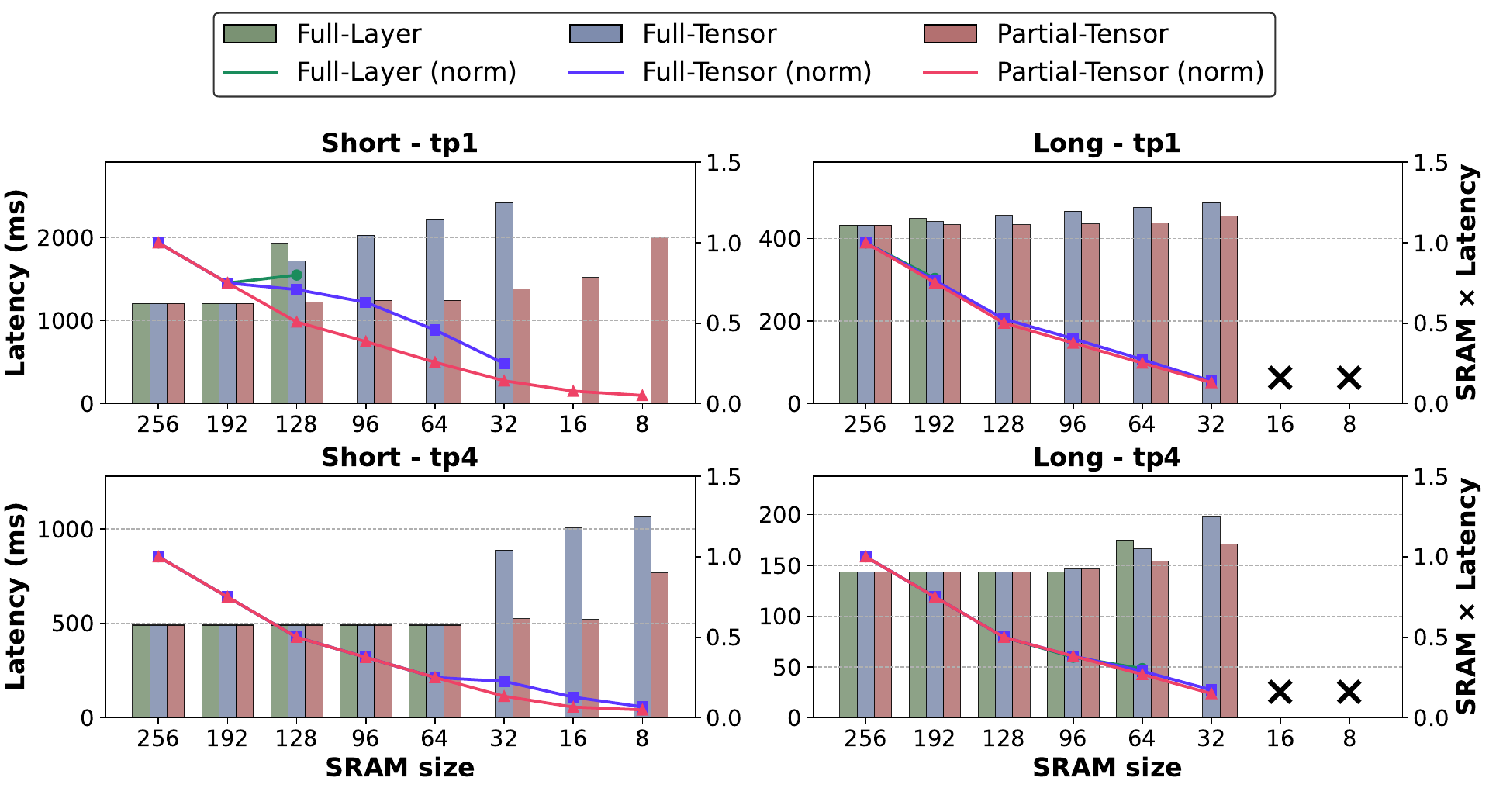}
    \caption{Latency of single-request execution under different tensor loading strategy,
    TP size and prefill size (short for MultiTurn Chat, long for TriviaQA). PP size is fixed to 2, and DP size is automatically determined by the compute-core count.}
    \label{fig:sram_size}
\end{figure}

To assess the impact of SRAM capacity on model inference, we evaluate three tensor-loading strategies:
(1) loading the full layer weights at once (full-layer),
(2) loading only the weights needed for each compute operation (full-tensor), and
(3) incrementally loading small weight fragments (partial-tensor, as mentioned in \textsection\ref{sub:design:mem}).
Fig.~\ref{fig:sram_size} shows the end-to-end latency for Qwen3\_4B on a 64-compute-core configuration with PP=2 and TP in \{1,4\}.

For both full-layer and full-tensor models, latency remains low as long as the SRAM is sufficiently large.
When SRAM falls below a critical threshold, latency degrades sharply because the full model weights can no longer fit into SRAM.
This threshold decreases as TP increases; for example, 128 MB at TP = 1 and 32 MB at TP = 4.

In contrast, partial-tensor shows minimal sensitivity to SRAM size and can operate efficiently with smaller SRAM budgets,
delivering up to a 2$\times$ speedup in short-prefill workloads (MultiTurn Chat).
In long-prefill scenarios (TriviaQA), however, prefill time dominates overall latency,
leaving little performance difference among the three strategies.

From these results, we observe that under the partial-tensor strategy, varying the SRAM capacity has a smaller impact on end-to-end latency.
This finding suggests that when weight access is sufficiently fine-grained,
the system becomes far less dependent on large on-chip memory,
enabling more flexible hardware configurations and reducing the need for aggressive SRAM provisioning.

\subsection{Different PD Strategies for LLM Serving}
Experiments in this section apply the previously selected strategy that best suits the corresponding scenario.

\begin{figure}[ht]
    \setlength{\abovecaptionskip}{-1pt}
    \setlength{\belowcaptionskip}{-15pt}
    \includegraphics[width=\linewidth]{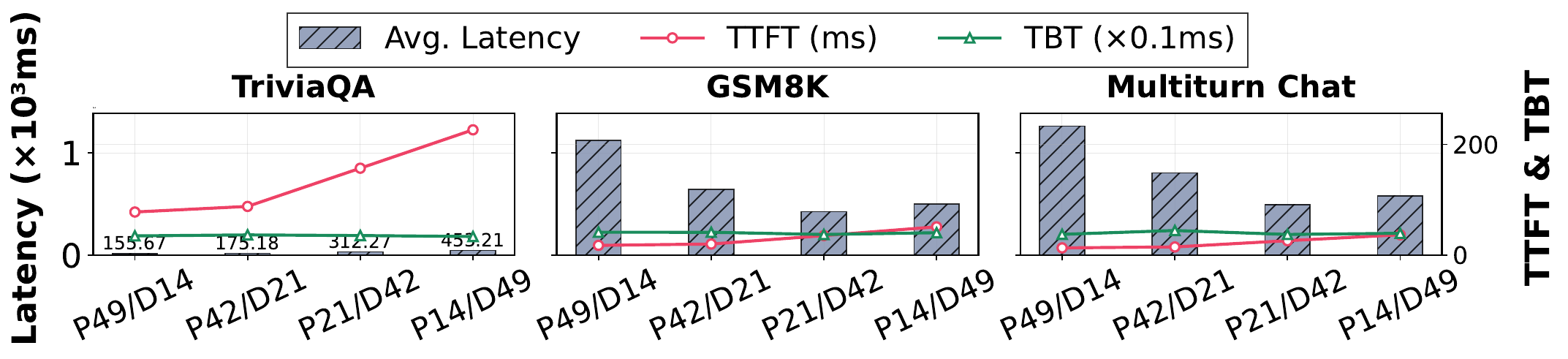}
    \caption{The effect of prefill-decoding core ratio on LLM serving performance. The degrees of TP and PP are kept constant, while the total compute-core count allocated to prefill and decoding is adjusted by varying the DP size.}
    \label{fig:pd_core_allocation}
\end{figure}

\myparagraph{Different core ratios in PD disaggregation:}
Fig.~\ref{fig:pd_core_allocation} shows the effect of varying prefill-decoding core ratios on LLM serving performance under 3 different workloads
(TriviaQA~\cite{TriviaQA}, GSM8K~\cite{GSM8K}, MultiTurn Chat~\cite{MultiturnChat}), as mentioned above.
In this evaluation, we take Qwen3\_4B with a 64-compute-core configuration as an example.

In prefill-dominated workloads, increasing prefill cores consistently reduces TTFT (Time to First Token),
while TBT (Time Between Tokens) remains largely insensitive to core allocation.
A larger prefill-core allocation therefore yields substantial benefits:
under the TriviaQA workload, P42/D21 provides about 3$\times$ TTFT speedup over P14/D49.
Although P49/D14 achieves the best TTFT, its advantage over P42/D21 is marginal,
making P42/D21 a more practical choice when considering overall system efficiency.

In contrast, for PD-balanced or decoding-dominated workloads, increasing the decoding compute-core count can reduce end-to-end latency by up to 2.6x,
and the latency exhibits a roughly U-shaped trend.
In this case, the P21/D42 configuration is preferred,
as it minimizes end-to-end latency by up to 49\% with a moderate TTFT trade-off.
However, further allocating cores to the decoding stage causes the prefill to become the bottleneck again, leading to increased latency.

In summary, the core-allocation ratio should be chosen according to workload characteristics.
Specifically, a P:D ratio of 2:1 is preferred for PD-dominated workloads,
while a P:D ratio of 1:2 is preferred for PD-balanced and decode-dominated workloads.
Adopting extreme allocations may flip the dominant execution phase (from prefill- to decode-dominated, or vice versa),
ultimately degrading system performance.

\begin{figure}[htp]
    \setlength{\abovecaptionskip}{-1pt}
    \setlength{\belowcaptionskip}{-18pt}
    \includegraphics[width=\linewidth]{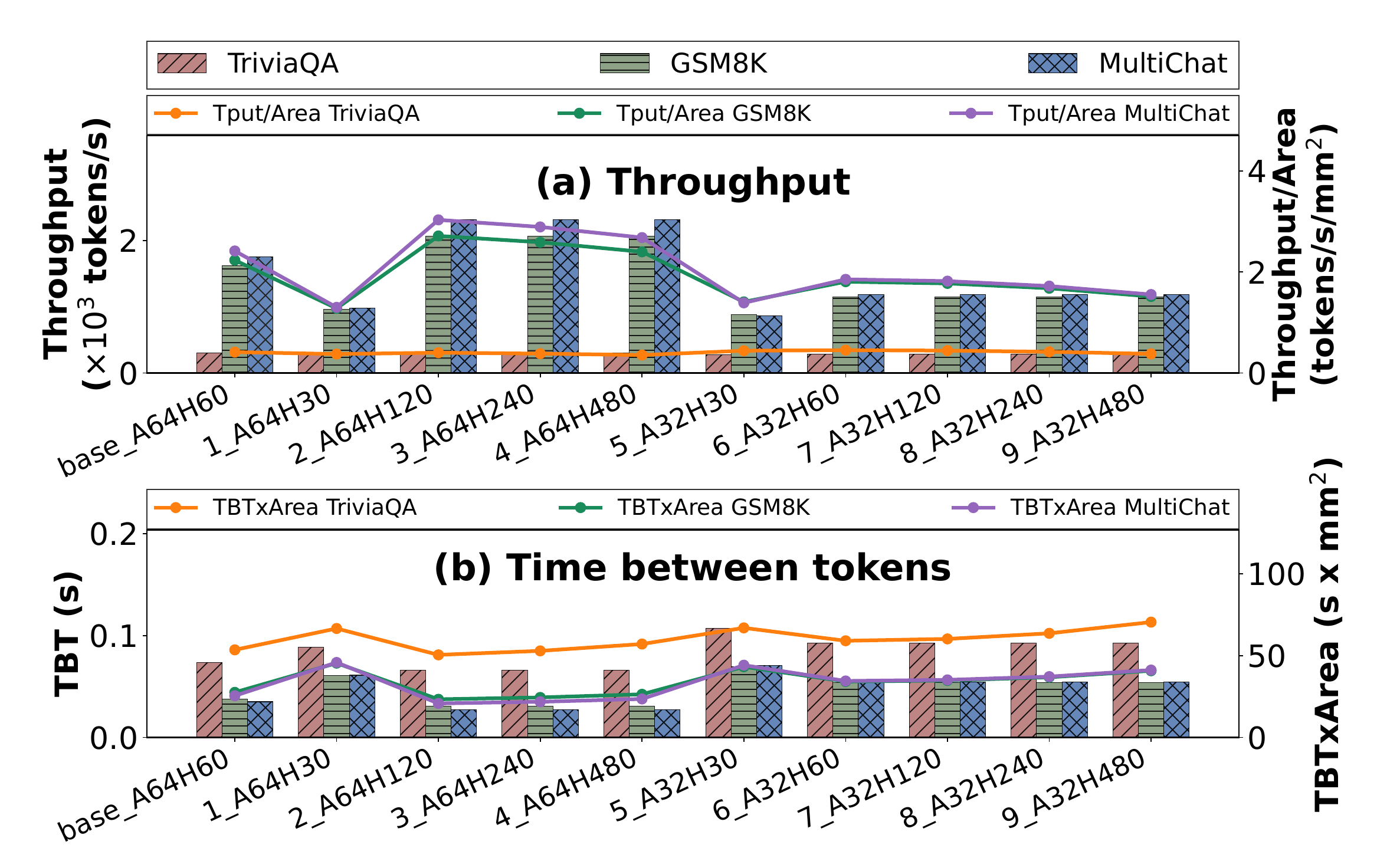}
    \caption{\textbf{The effects of different hardware configurations on serving throughput, TBT for heterogeneous PD disaggregation scenarios.}
    The X-axis represents different configurations for the decoding core: ``A'' denotes the dimension of the systolic array,
    and ``H'' denotes the per-core HBM bandwidth (in GB/s).}
    \label{fig:pd_hete_analysis}
\end{figure}

\myparagraph{Heterogeneous core design for PD disaggregation:}
We investigate heterogeneous resource allocation for prefill and decode by varying two key architectural parameters: systolic-array dimensions and HBM bandwidth.
Since prefill is compute-intensive and decoding is memory-intensive, we scale SRAM bandwidth proportionally with systolic array's compute capacity.
All experiments adopt a P/D core allocation ratio of 2:1.
For the area-normalized results, we use a component-level die-shot fitting model rather than an abstract area formula. The modeled components are compute cubes, on-chip SRAM blocks, and HBM interfaces, whose area and bandwidth are estimated from the Ascend 910B die shot.\footnote{In the fitted die-shot model, the area of one compute cube is estimated as 6.51\,mm$^2$, including a $16\times16\times16$ cube, vector units, and other local logic; the area of one 8\,MB SRAM block is estimated as 3.24\,mm$^2$; and the area of one 800\,GB/s HBM interface is estimated as 12.28\,mm$^2$, with the controller and physical-interface parts estimated as 3.36\,mm$^2$ and 6.72\,mm$^2$, respectively.}
The area denominator for each configuration is obtained by scaling these fitted component units according to the configured systolic-array dimension, SRAM capacity, and per-core HBM-interface bandwidth. Across the ten configurations in Fig.~\ref{fig:pd_hete_analysis}, the resulting total estimated chip area ranges from 623.9 to 861.5\,mm$^2$.

As shown in Fig.~\ref{fig:pd_hete_analysis}(a), increasing the decoding-core HBM bandwidth improves throughput by up to 1.31x,
and area-normalized throughput by 1.25x (Configuration 2).
Beyond this point, further bandwidth increases (Configurations 3-4) yield no additional benefit,
indicating a shift of the performance bottleneck from memory to compute. TBT exhibits a similar trend shown in Fig.~\ref{fig:pd_hete_analysis}(b), where increasing the decoding-core bandwidth consistently
improves area-normalized TBT.

However, reducing compute capacity results in 16.7\% worse throughput and TBT than the baseline.
This is because lower compute capability reduces the efficiency of GEMV operations in the decode phase,
increasing its latency and degrading overall performance.
Moreover, when prefill dominates the execution(e.g., TriviaQA workloads), both throughput and TBT become insensitive to heterogeneous decode-core configurations,
as decoding contributes only a small fraction of end-to-end latency and low token counts limit compute utilization.

\begin{figure}[ht]
    \setlength{\abovecaptionskip}{-1pt}
    \setlength{\belowcaptionskip}{-10pt}
    \includegraphics[width=\linewidth]{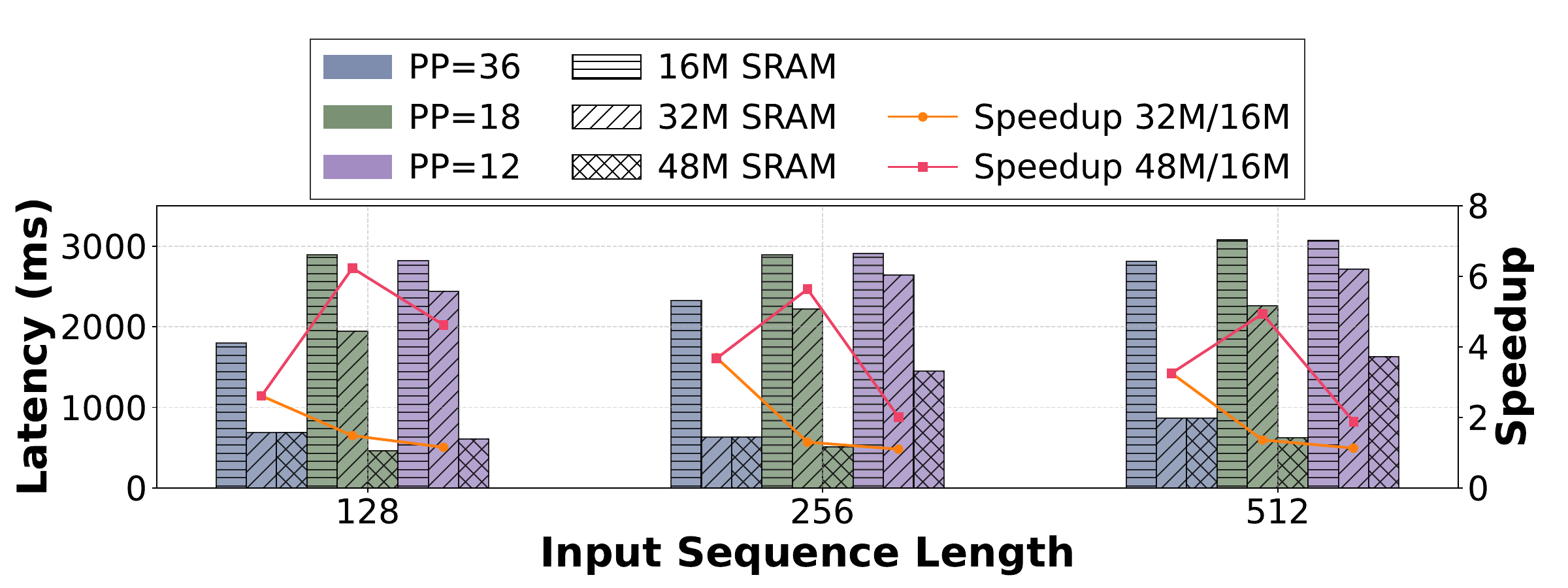}
    \caption{End-to-end latency of PD fusion with different input token lengths, per-core SRAM sizes, and pipeline stage counts for Qwen3\_8B (TP = 4) on a 256-compute-core configuration.}
    \label{fig:pd_fusion_pipeline}
\end{figure}

\myparagraph{Hardware optimization under PD fusion:}
Fig.~\ref{fig:pd_fusion_pipeline} presents the impact of input token length, SRAM capacity, and pipeline stage count on end-to-end latency under PD fusion.

With fewer pipeline stages, each core processes more layers, thereby increasing data parallelism.
However, this also increases the memory pressure on each core, resulting in more frequent SRAM spilling.
Therefore, with a small SRAM size (16MB), 36 pipeline stages achieve 1.1x-1.61x performance improvement compared to the 18 and 12 pipeline stages, respectively.

As the memory pressure increases due to the PD fusion design,
increasing the SRAM capacity leads to more significant improvements in inference performance.
For example, expanding the SRAM from 16 MB to 32 MB results in a 2.6x-3.7x performance speedup.
In addition, a larger SRAM capacity can support fewer pipeline stages without causing data spill,
thereby achieving improved performance.

\begin{figure}[ht]
    \setlength{\abovecaptionskip}{0pt}
    \setlength{\belowcaptionskip}{-10pt}
    \includegraphics[width=\linewidth]{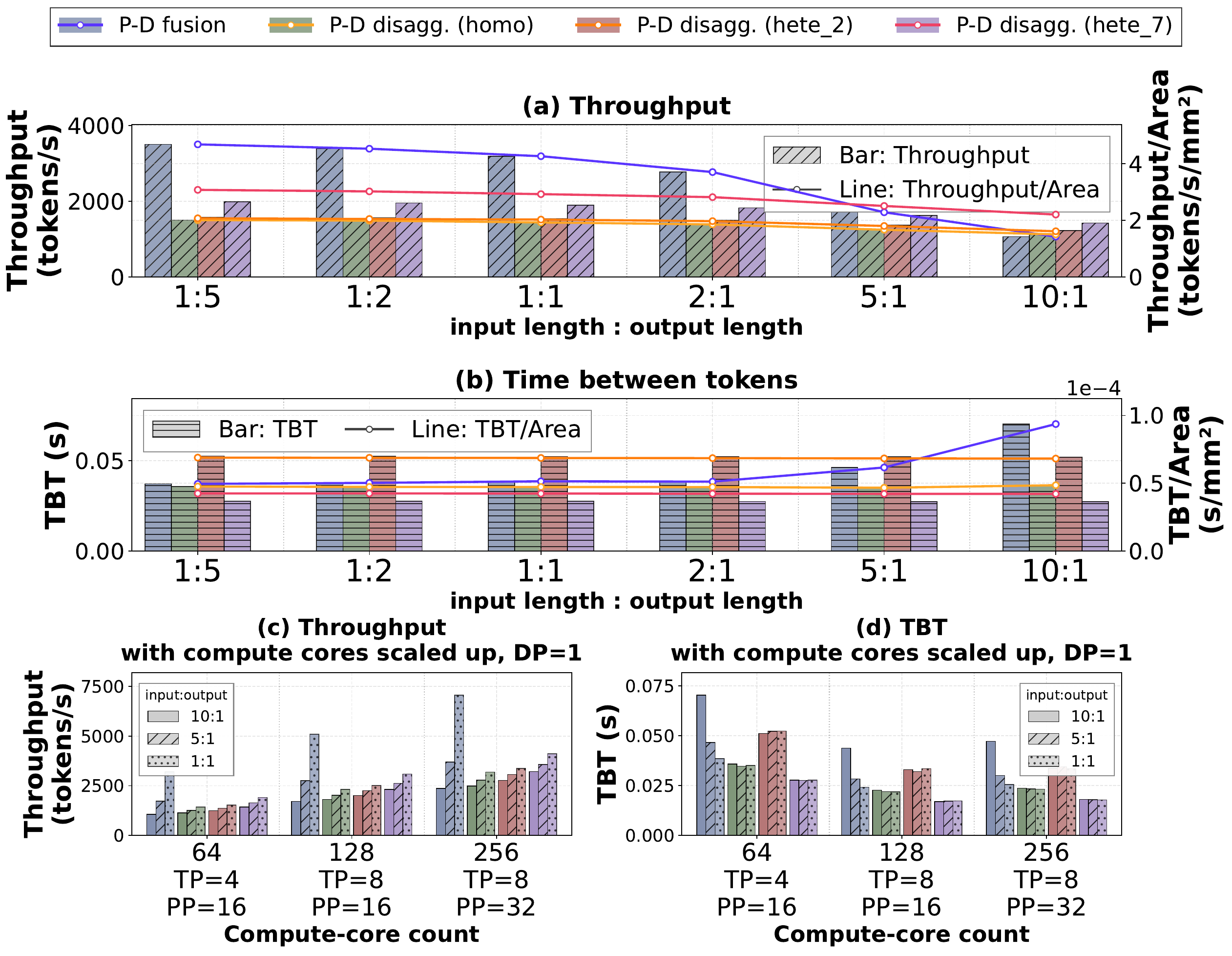}
    \caption{Throughput and TBT comparison between PD disaggregation and PD fusion under varying prefill-to-decode length (input:output tokens) ratios
    for Qwen3\_4B. Subplots~(c) and~(d) show scale-up results for P:D ratios of 10:1, 5:1, and 1:1 at compute-core counts of 64, 128, and 256.}
    \label{fig:split_vs_fuse}
\end{figure}

\myparagraph{Comparison of PD disaggregation and PD fusion:}
Fig.~\ref{fig:split_vs_fuse} compares PD disaggregation and PD fusion in throughput and TBT across workloads, using two heterogeneous configurations and one homogeneous baseline from Fig.~\ref{fig:pd_hete_analysis} to highlight heterogeneous PD disaggregation.
On homogeneous cores, PD fusion gives 6\% higher throughput and lower TBT than PD disaggregation in decode-dominated workloads; in prefill-dominated workloads, however, decoding interferes with prefill, increases TTFT, and reduces PD fusion's effectiveness, causing a 28\% throughput decline relative to PD disaggregation.
Across configurations, PD disaggregation with heterogeneous decode cores outperforms PD fusion in most scenarios, reaching up to 37\% higher throughput and lower TBT; thus, when heterogeneity is properly configured for real workloads, PD disaggregation can offer superior efficiency.
The scale-up results in Fig.~\ref{fig:split_vs_fuse}(c) and Fig.~\ref{fig:split_vs_fuse}(d) show that the trend remains under the evaluated compute-core counts of 64, 128, and 256 for single-instance scaling, although the preferred prefill/decode allocation ratio remains workload-dependent.

\section{CONCLUSION}
In this paper, we systematically analyze the hardware architecture design of multi-core AI accelerators and optimization strategies for LLM serving scenarios.
Leveraging our efficient and configurable multi-core AI accelerator simulator,
we explore various hardware configuration strategies, tensor parallelism and placement methods, hybrid memory management techniques,
as well as heterogeneous design for PD stages.
Experimental results demonstrate that the evaluated strategies consistently reduce latency across core-placement and tensor-partitioning studies, showing the benefit of architecture-aware co-design for LLM serving on multi-core AI accelerators.
We hope that our findings will inspire further architectural innovations and system-level optimizations for multi-core AI accelerators in LLM serving.
\label{s:conclusion}

\appendix
\vspace{0.5em}

This appendix provides the requested reproducibility and interpretation materials: an isolated K-dimension communication experiment, the Matmul compute--memory estimator, and PD serving schedulers.

\begin{figure}[H]
    \centering
    \includegraphics[width=\linewidth]{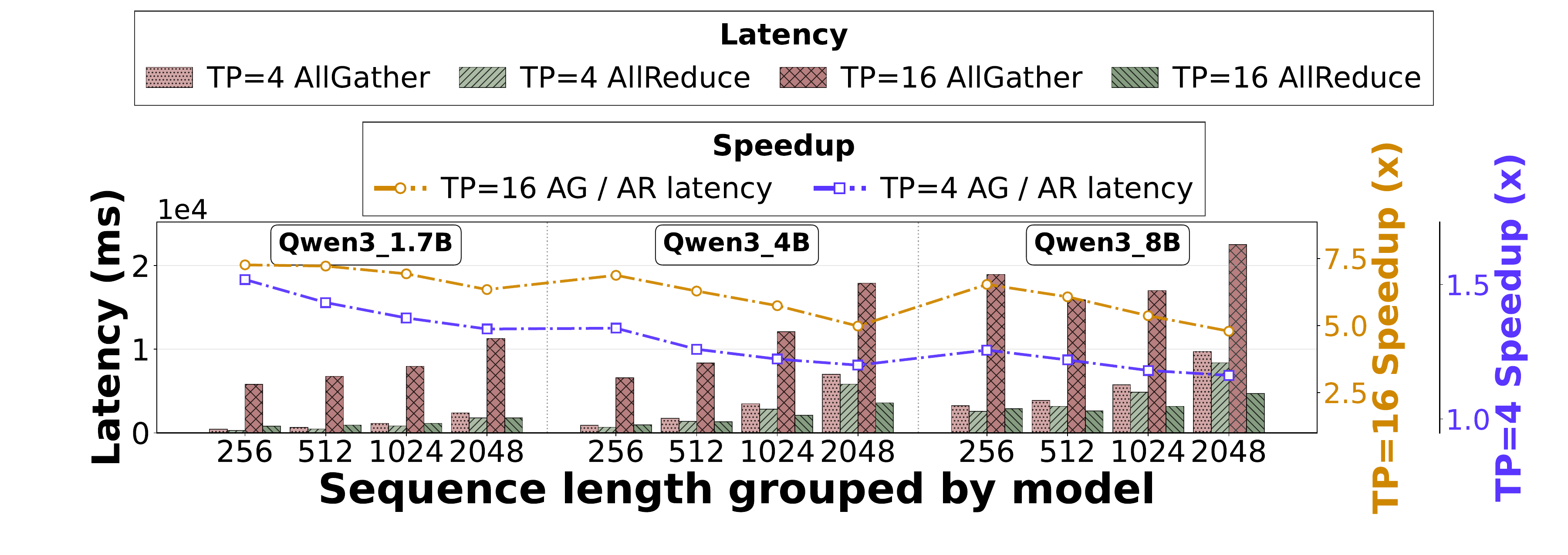}
    \caption{Isolated AllGather-vs.-AllReduce comparison for K-dimension partition at TP=4/16, varying only the collective primitive while fixing model, sequence length, placement, memory strategy, and scheduling. Bars: latency; lines: ratios.}
    \label{fig:tp_kdim_comm}
\end{figure}

Fig.~\ref{fig:tp_kdim_comm} isolates the post-K-partition collective: AllReduce is consistently faster than AllGather, and its gap grows from TP=4 to TP=16, matching its larger communication-volume advantage at higher TP degrees.

Algorithm~\ref{alg:app-matmul} summarizes \sys's Matmul latency model: separate analytical compute and transaction-level memory timing, then merge them with compute--memory overlap.

\begin{algorithm}[H]
\caption{Matmul compute/memory latency estimation.}
\label{alg:app-matmul}
\label{alg:app-matmul-compute}
\label{alg:app-matmul-memory}
{\footnotesize
\begin{algorithmic}[1]
\STATE \textbf{GEMM:} $(M_g \times K_g) \times (K_g \times N_g)$, batch $B$.
\STATE \textbf{HW:} systolic $S$; EXUs $C_{\text{exu}}$; vector $V{\times}C_{\text{vec}}$; SFU throughput $P_{\text{sfu}}$ if supported.
\STATE \hrulefill
\STATE \textbf{Compute latency:}
\STATE $N_{\text{tiles}} \leftarrow \lceil K_g / S \rceil \cdot \lceil N_g / S \rceil$
\STATE $M_{\mathrm{pad}} \leftarrow \max(B \cdot M_g,\, S)$
\STATE $T_{\text{cycles}} \leftarrow M_{\mathrm{pad}} + 2S - 2$;\quad $T_{\text{inject}} \leftarrow 2S$
\STATE $\tau_{\text{exu}} \leftarrow \lceil N_{\text{tiles}} / C_{\text{exu}} \rceil \cdot T_{\text{cycles}} + T_{\text{inject}}$
\STATE $O_{\text{vec}} \leftarrow 2 \cdot B \cdot M_g \cdot K_g \cdot N_g$;\quad $P_{\text{vec}} \leftarrow V \cdot C_{\text{vec}}$
\STATE $\tau_{\text{vec}} \leftarrow O_{\text{vec}} / P_{\text{vec}}$
\STATE $\tau_{\text{sfu}} \leftarrow \infty$ if Matmul unsupported on SFU, else $O_{\text{sfu}} / P_{\text{sfu}}$
\STATE $\tau_{\text{comp}} \leftarrow \min(\tau_{\text{exu}}, \tau_{\text{vec}}, \tau_{\text{sfu}})$
\STATE \hrulefill
\STATE \textbf{Memory latency:} buffer map stores offset, size, and residency/spill per tensor.
\STATE $T_{\text{mem}} \leftarrow 0$
\FOR{each static tensor (weight/bias)}
\IF{not resident \textbf{or} spilled}
\STATE Fetch from DRAM; update map; $T_{\text{mem}} \mathrel{+}=$ transaction time
\ENDIF
\ENDFOR
\FOR{$i = 1$ to $\lceil N_g/S \rceil - 1$}
\IF{activation is off-chip}
\STATE Read slab; $T_{\text{mem}} \mathrel{+}=$ transaction time
\ENDIF
\ENDFOR
\STATE $\tau_{\text{eff}} \leftarrow \max(T_{\text{mem}}, \tau_{\text{comp}})$
\end{algorithmic}
}
\end{algorithm}

Algorithm~\ref{alg:app-pd-schedulers} summarizes the two PD scheduling modes: fusion and disaggregation are selected simulation modes, not dynamically switched states; prefill-to-decoding movement is a request-state handoff within the selected scheduler.

\begin{algorithm}[H]
\caption{PD fusion/disaggregation scheduling.}
\label{alg:app-pd-schedulers}
\label{alg:app-pd-fusion-scheduler}
\label{alg:app-pd-disaggregation-scheduler}
{\footnotesize
\begin{algorithmic}[1]
\STATE \textbf{Fusion:} queues $Q_R,Q_P,Q_D,\mathrm{CD}$; budget $C$; prefill cost $R$.
\STATE \textbf{if} group busy \textbf{return}; propagate pipeline.
\FOR{each first-stage core}
\STATE $B{\leftarrow}C$; take $\mathrm{CD}$ while $B>0$; spill overflow to $Q_D$
    \WHILE{$B>0$}
        \IF{$Q_D \neq \emptyset$}
\STATE issue $\mathrm{Decode}(\mathrm{pop}(Q_D))$; $B \leftarrow B-1$
\ELSIF{$B \ge R$ \textbf{and} ($Q_P \neq \emptyset$ \textbf{or} $Q_R \neq \emptyset$)}
\STATE $r{\leftarrow}\mathrm{pop}(Q_P)$ else $\mathrm{pop}(Q_R)$; issue $\mathrm{PrefillChunk}(r)$
\STATE $B \leftarrow B-R$; requeue $r$ if chunks remain
        \ELSE
            \STATE \textbf{break}
        \ENDIF
    \ENDWHILE
\ENDFOR
\STATE Issue configs; set busy.
\STATE \hrulefill
\STATE \textbf{Disagg:} groups $G_P,G_D$; queues $Q_R,Q_P,Q_{P\to D},Q_D,\mathrm{CD}$.
\STATE \textbf{Input:} $m{\in}\{P,D\}$; \textbf{if} $G_m$ busy \textbf{return}; propagate $G_m$.
\IF{$m = P$}
    \FOR{each prefill first-stage core}
\STATE continue $Q_P$; admit $Q_R$ up to $\mathrm{batch\_size}$; requeue leftovers
    \ENDFOR
\ELSIF{$m = D$}
    \FOR{each decode first-stage core}
\STATE $B{\leftarrow}C$; take $\mathrm{CD}$; drain $Q_D,Q_{P\to D}$ while $B>0$
    \ENDFOR
\ENDIF
\STATE Issue configs for $G_m$; update busy.
\end{algorithmic}
}
\end{algorithm}

\section*{Acknowledgment}
We sincerely thank our anonymous reviewers, area editor Liu Weichen, associate editor Costero Luis for their insightful suggestions. This work is supported in part by China National Natural Science Foundation (No. 62472279). Erhu Feng
is the corresponding author.
The authors disclose that OpenAI Codex was used only to assist with language editing and polishing of the manuscript text.
The research design, overall paper organization, experimental implementation, testing, and evaluation were performed by the authors.
The authors reviewed and take full responsibility for all content in this manuscript.

\bibliographystyle{IEEEtran}
\bibliography{references}

\end{document}